\documentclass[]{aastex631}
\usepackage{amsmath}

\received{}
\revised{}
\accepted{}

\submitjournal{The Astrophysical Journal}

\shorttitle{A Simulation Study of Ultra-Relativistic Jets}
\shortauthors{Seo et al.}

\begin{document}

\title{A Simulation Study of Ultra-relativistic Jets - II. Structures and Dynamics of FR-II Jets}
\author[0000-0002-5550-8667]{Jeongbhin Seo}
\affiliation{Department of Earth Sciences, Pusan National University, Busan 46241, Korea}
\author[0000-0002-4674-5687]{Hyesung Kang}
\affiliation{Department of Earth Sciences, Pusan National University, Busan 46241, Korea}
\author[0000-0002-5455-2957]{Dongsu Ryu}
\affiliation{Department of Physics, College of Natural Sciences, UNIST, Ulsan 44919, Korea}
\correspondingauthor{Dongsu Ryu}\email{ryu@sirius.unist.ac.kr}

\begin{abstract}


We study the structures of ultra-relativistic jets injected into the intracluster medium (ICM) and the associated flow dynamics, such as shocks, velocity shear, and turbulence, through three-dimensional relativistic hydrodynamic (RHD) simulations. To that end, we have developed a high-order accurate RHD code, equipped with a weighted essentially non-oscillatory (WENO) scheme and a realistic equation of state \citep[][Paper I]{seo2021a}. Using the code, we explore a set of jet models with the parameters relevant to FR-II radio galaxies. We confirm that the overall jet morphology is primarily determined by the jet power, and the jet-to-background density and pressure ratios play secondary roles. Jets with higher powers propagate faster, resulting in more elongated structures, while those with lower powers produce more extended cocoons. Shear interfaces in the jet are dynamically unstable, and hence, chaotic structures with shocks and turbulence develop. We find that the fraction of the jet-injected energy dissipated through shocks and turbulence is greater in less powerful jets, although the actual amount of the dissipated energy is larger in more powerful jets. In lower power jets, the backflow is dominant in the energy dissipation owing to the broad cocoon filled with shocks and turbulence. In higher power jets, by contrast, both the backflow and jet spine flow are important for the energy dissipation. Our results imply that different mechanisms, such as diffusive shock acceleration, shear acceleration, and stochastic turbulent acceleration, may be involved in the production of ultra-high energy cosmic rays in FR-II radio galaxies.

\end{abstract}


\section{Introduction}

Radio jets, driven by active galactic nuclei (AGNs), can expand out and inflate X-ray cavities of up to $\sim$ Mpc scales in the intracluster medium (ICM) of galaxy clusters \citep[see, e.g.,][for reviews]{begelman1984,fabian2012}. They could significantly disturb the thermal and dynamic properties of the ICM through the injection of heat, relativistic particles, shock waves, turbulence, and magnetic fields. Conventionally, extended radio galaxies are classified into two distinct morphological types, the center-brightened FR-I with a pair of plums and the edge-brightened FR-II with a pair of hot spots \citep{fanaroff1974}. It is thought that the jets of FR-I are decelerated to sub-relativistic speeds on kpc scales through the entrainment of ambient media and the mass-loading by stellar winds, as well as the dissipation due to small-scale instabilities \citep[e.g.,][]{bicknell1984,komissarov1994,laing2014,perucho2014,perucho2019,rossi2020}. On the other hand, the FR-II jets remain relativistic up to $\sim 100$ kpc until they are halted at the jet head
\citep[e.g.,][]{bicknell1995,tchekhovskoy2016,hardcastle2018}.

The FR-I and FR-II dichotomy is thought to originate primarily from the difference in the jet power, $Q_j$ (see Eq. [\ref{Qjet}]) \citep[e.g.,][]{kaiser1997,godfrey2013}, which is determined mostly by the initial speed, $v_j/c$, or the initial bulk Lorentz factor of the jet, $\Gamma_j=1/\sqrt{1-(v_j/c)^2}$, for given jet radius and density. In addition, the jet-to-background density contrast, $\eta\equiv\rho_j/\rho_b$, and pressure contrast, $\zeta\equiv p_j/p_b$, are the secondary parameters for the dichotomy \citep{norman1982,hardcastle2018,rossi2020}; alternatively, the momentum injection rate, or the jet thrust, $\dot{M}_j$, (see Eq. [\ref{Mjet}]) may be used to describe the dichotomy \citep[e.g.,][]{perucho2014,hardcastle2020}. Hereafter, the subscripts ``$j$'' and ``$b$'' denote the states of the jet material and the background medium, respectively. 

According to the analysis of observed radio luminosities by \citet{godfrey2013}, FR-I radio galaxies are inferred to be driven by less powerful jets with $Q_j\lesssim 10^{45}{\rm erg~s^{-1}}$, while FR-II radio galaxies are to be induced by more powerful jets with $Q_j\gtrsim 10^{45}{\rm erg~s^{-1}}$. Both the types are found in the intermediate range of $Q_j\sim 10^{45}-10^{46}{\rm erg~s^{-1}}$. The existence of this overlapping range can be understood naturally, since the jet dynamics depends not only on $Q_j$, but also on $\eta$ and $\zeta$. Through extensive simulations, \citet{li2018} showed that the jets with slower speeds (smaller $\Gamma_j$) and lower densities (smaller $\eta$) tend to produce the unstable FR-I type morphology, while those with faster speeds (larger $\Gamma_j$) and higher densities (larger $\eta$) induce the stable FR-II type morphology (see their Figure 1). However, for FR-I jets, it is still challenging to precisely constrain $Q_j$ from radio and X-ray observations or to numerically simulate the dynamical evolution with realistic treatments of entrainment and dissipative processes \citep[e.g.][]{perucho2019,perucho2020}.

The morphology and also the dynamics of relativistic jets were studied  by both analytical modelings \citep[e.g.,][]{scheuer1974,blandford1974,kaiser1997} and hydrodynamic (HD) simulations \citep[e.g.,][]{norman1982, reynolds2002, krause2005, hardcastle2013}. More recently, numerical studies have been expanded to include the magnetic field and special relativistic effects through magnetohydrodynamic (MHD) simulations \citep[e.g.,][]{tregillis2001,oneill2005,mendygral2012,hardcastle2014}, relativistic hydrodynamic (RHD) simulations \citep[e.g.,][]{perucho2007,perucho2019,english2016,li2018}, and relativistic magnetohydrodynamic (RMHD) simulations \citep[e.g.,][]{leismann2005,porth2015,marti2016,tchekhovskoy2016}. A comprehensive review on numerical studies of AGN jets can be found in \citet{marti2019}.
 
Giant radio galaxies are thought to be possible sites for the production of ultra-high energy cosmic rays (UHECRs) \citep[see][for recent reviews]{blandford2019,rieger2019,hardcastle2020,matthews2020}. Through interactions with the ambient medium, AGN jets induce complex flows with shocks, velocity shear, and turbulence, and UHECRs could be accelerated there. In addition, the magnetic field carried by the jet plasmas can be amplified by the flow motions. With the characteristic size of radio lobes up to $\sim100$ kpc and the magnetic field up to $\sim 100\ \mu$G, giant radio galaxies can accommodate the UHECR protons of the energy up to the order of 100 EeV \citep{hillas1984}.

Possible acceleration mechanisms of UHECRs, which have been suggested so far, include the first-order Fermi (Fermi-I) acceleration (diffusive shock acceleration) mainly at sub-relativistic shocks in the jet-induced backflow \citep{matthews2019}, the stochastic, second-order Fermi (Fermi-II) acceleration by turbulent flows in the lobe \citep{hardcastle2010}, the gradual shear acceleration in relativistic shearing flows \citep{rieger2004,rieger2019,webb2018}, the discrete (non-gradual) shear acceleration at the interface between the jet spine and backflow \citep{ostrowski1998,kimura2018}, and the espresso mechanism with one or a few shot boosts by the ultra-relativistic jets of $\Gamma \sim 10-30$ \citep{caprioli2015,mbarek2019}. On the other hand, the particle acceleration by relativistic magnetic reconnection could be important in the compact, ultra-relativistic jets of gamma-ray bursts and blazars with strong magnetic fields \citep{sironi2014,petropoulou2019}.

In this paper, through RHD simulations, we study the structures and flow dynamics of ultra-relativistic jets with the parameters relevant to FR-II radio galaxies.\footnote{Here, we do not consider FR-I jets with $Q_j\lesssim 10^{45}{\rm erg~s^{-1}}$. Modeling of realistic FR-I jets may need to include the entrainment and mass-loading on kpc scales and the dissipation through small-scale instabilities. We leave the study of FR-I jets as a future work.} Especially, we examine and quantify the distributions and properties of shocks, velocity shear, and turbulence produced by jets, and then estimate the amount of the jet energy dissipated at different regions of jet-induced structures. Aiming to follow the nonlinear flows with high accuracy and robustness, we use a newly developed three-dimensional (3D) RHD code based on the 5th-order accurate, finite-difference WENO (weighted essentially non-oscillatory) scheme for solving hyperbolic conservation equations \citep{jiang1996,borges2008} and the 4th-order accurate, strong stability preserving Runge–Kutta (SSPRK) time discretization \citep{spiteri2003}. In addition, to correctly reproduce the thermodynamics across the jet and the ICM, the code incorporates the RC version of equation of state (EOS), which closely approximates the EOS of the perfect gas in relativistic regime \citep{ryu2006}. The details of the new RHD code including tests to demonstrate the performance can be found in the companion paper \citep[][hereafter Paper I]{seo2021a}.

We point that the presence of $\sim10-100$ $\mu$G magnetic fields in radio galaxies has been established, for instance, through the analysis of synchrotron emission \citep[e.g.,][]{heinz1997} or the equipartition estimate \citep[e.g.,][]{godfrey2013}. On the other hand, some observations hint that the jet evolution on kpc and larger scales would be primarily governed by the jet kinetic energy power \citep[e.g.,][]{rawlings1991}. Moreover, simulation studies indicate that the magnetic field may not be dynamically crucial in reproducing the observed morphology \citep[e.g.,][]{clarke1986}. Yet, the magnetic field could play important roles in determining the flow dynamics on small scales. In this study, we do not include the magnetic field, concentrating on the RHDs of relativistic jets.

The paper is organized as follows. In Section \ref{s2} the details of simulations are described. The results of simulations, that is, the morphology and dynamics of jets, are presented in Sections \ref{s3} and \ref{s4}. A brief summary follows in Section \ref{s5}.

\section{Jet Simulations}
\label{s2}

\subsection{Basic Equations}
\label{s2.1}

The conservation equations for special RHDs can be written as
\begin{eqnarray}
\frac{\partial D}{\partial t} + \mbox{\boldmath$\nabla$}\cdot(D\mbox{\boldmath$v$}) = 0, \label{Deq}\\
\frac{\partial \mbox{\boldmath$M$}}{\partial t} + \mbox{\boldmath$\nabla$}\cdot(\mbox{\boldmath$M$}\mbox{\boldmath$v$}+p) = 0, \label{Meq}\\
\frac{\partial E}{\partial t} + \mbox{\boldmath$\nabla$}\cdot[(E+p)\mbox{\boldmath$v$}] = 0\label{Eeq}
\end{eqnarray}
\citep[e.g.,][]{landau1959}. The conserved quantities, $D$, $\mbox{\boldmath$M$}$, and $E$, are the mass, momentum, and total energy densities in the laboratory frame, respectively, which are related to the primitive variables, the rest-mass density, $\rho$, the fluid three-velocity, $\mbox{\boldmath$v$}$, and the isotropic gas pressure, $p$, as
\begin{eqnarray}
D = \Gamma\rho, \label{Dcons}\\
\mbox{\boldmath$M$} = \Gamma^{2}\rho\frac{h}{c^2} \mbox{\boldmath$v$}, \label{Mcons}\\
E = \Gamma^{2}\rho h -p. \label{Econs}
\end{eqnarray}
Here, $c$ is the speed of light, $\Gamma=1/\sqrt{1-(v/c)^{2}}$ with $v=|\mbox{\boldmath$v$}|$ is the Lorentz factor, $h = (e + p)/\rho$ is the specific enthalpy, and $e$ is the sum of the internal and rest-mass energy densities.

We adopt the EOS proposed by \citet{ryu2006}, which was named as the RC EOS in Paper I:
\begin{equation}
\frac{h}{c^2} = 2\frac{6\Theta^{2}+4\Theta+1}{3\Theta+2},
\label{hRCEOS}
\end{equation}
where $\Theta\equiv p/\rho c^2$ is a temperature-like variable. It closely approximates the EOS of the single-component perfect gas in relativistic regime, called the RP EOS in Paper I. Hence, our RC EOS correctly describes the fluids of both non-relativistic temperature ($\Theta \lesssim 1$) and relativistic temperature ($\Theta \gtrsim 1$).

Although the adiabatic index, $\gamma$, does not explicitly appear in the RHD equations, below we present $\gamma$ which is estimated with $\rho$ and $p$ from
\begin{equation}
\frac{\gamma}{\gamma-1} = \frac{h/c^2-1}{\Theta}.
\label{hIDEOS}
\end{equation} 
It ranges between $\gamma=5/3$ for $\Theta\ll1$ and $4/3$ for $\Theta\gg1$.

\subsection{Background Medium}
\label{s2.2}

The domain of our simulations is a rectangular box in the 3D Cartesian coordinate system, whose bottom surface defined by $z=0$ contains a circular jet nozzle with the radius of $r_j=1$~kpc, centered at $(x,y)=(0,0)$. At the onset of simulations, the box is filled with a static uniform background medium with the density, $\rho_b$, and the pressure, $p_b$. Without including dissipative processes such as radiative losses, the RHD equations in (\ref{Deq})-(\ref{Eeq}) are scalable for arbitrary length, time, and density, so are our simulations. However, we adopt the typical parameters for the hot ICM, that is, $n_{\rm H,ICM} \approx 10^{-3}$ cm$^{-3}$ for the hydrogen number density and $T_{\rm ICM} \approx 5\times10^7$ K for the temperature, since we are interested in radio galaxies in galaxy clusters. Then, $\rho_b \approx 2.34\times 10^{-24} {\rm g} \cdot n_{\rm H,ICM}$ and $p_b\approx 2.3 \cdot n_{\rm H,ICM}k_{\rm B}T_{\rm ICM}$, where $k_{\rm B}$ is the Boltzmann constant. We do not consider the stratification of ICM halos, because the jets expand out only up to $\sim 50 - 60$ kpc in our simulations. Aiming to explore the dependence of jet structures and flow dynamics on jet parameters with a simple background configuration, our study focuses on the early evolution of radio galaxies on several tens of kpc scales in the ICM.

In the next sections, the results of our simulations are presented in units of $r_0=r_j=1$ kpc, $v_0=c$, and $\rho_0= \rho_b=2.34\times10^{-27} {\rm g~cm^{-3}}$ for the length, velocity, and density, respectively; the time and pressure are expressed in units of $t_0=r_j/c=3.26\times10^3~{\rm years}$ and $p_0=\rho_0 c^2=2.1\times10^{-6}~{\rm erg~cm^{-3}}$, respectively. Then, the pressure of the background medium is given as $p_b/p_0=7.64\times10^{-6}$ in the dimensionless unit, and its adiabatic index is $\gamma\approx 5/3$ with $p_b/\rho_bc^2\ll 1$.

\begin{deluxetable*}{lcccccccccccc}[t]
\tablecaption{Model Parameters\label{tab:t1}}
\tabletypesize{\small}
\tablecolumns{11}
\tablenum{1}
\tablewidth{0pt}
\tablehead{
\colhead{Model name} &
\colhead{$Q_j(\rm{erg~s^{-1}})$} &
\colhead{$\eta\equiv\frac{\rho_j}{\rho_b}$} &
\colhead{$\zeta\equiv\frac{p_j}{p_b}$} &
\colhead{$\dot{M}_j(\rm{dyne})$} &
\colhead{$v_{\rm j}/c$} &
\colhead{$\Gamma_{\rm j}$} &
\colhead{$v_{\rm head}^*/c$} &
\colhead{$t_{\rm cross}(\rm{yr})$} &
\colhead{Grid zones} &
\colhead{${N_j\equiv\frac{r_j}{\Delta x}}^a$} &
\colhead{${\frac{t_{\rm end}}{t_{\rm cross}}}^b$}
}
\startdata
Q45-$\eta5$-$\zeta0$ &3.34E+45&1.E-05&1&1.15E+35&0.9905&7.2644&0.0409 &2.66E+6&$(400)^2\times600$ & 12 & 48\\
Q46-$\eta5$-$\zeta0$ &3.34E+46&1.E-05&1&1.13E+36&0.9990&22.5645&0.1180 &9.22E+5&$(400)^2\times600$ & 12 & 37\\
Q46-$\eta5$-$\zeta0$-H &&&&&&&&&$(600)^2\times1200$ & 18 & 54\\
Q47-$\eta5$-$\zeta0$ &3.34E+47&1.E-05&1&1.12E+37&0.9999&71.0149&0.2965 &3.67E+5&$(400)^2\times600$ & 12 & 38\\
Q47-$\eta5$-$\zeta0$-H &&&&&&&&&$(600)^2\times1200$ & 18 & 50\\
\hline
Q45-$\eta4$-$\zeta0$ &3.34E+45&1.E-04&1&1.34E+35&0.9729&4.3259&0.0441 &2.47E+6&$(400)^2\times600$ & 12 & 50\\
Q45-$\eta3$-$\zeta0$ &3.34E+45&1.E-03&1&1.90E+35&0.8646&1.9899&0.0516 &2.11E+6&$(400)^2\times600$ & 12 & 49\\
Q46-$\eta4$-$\zeta0$ &3.34E+46&1.E-04&1&1.19E+36&0.9968&12.5690&0.1208&9.01E+5&$(400)^2\times600$ & 12 & 41\\
Q46-$\eta3$-$\zeta0$ &3.34E+46&1.E-03&1&1.37E+36&0.9774&4.7332&0.1283 &8.48E+5&$(400)^2\times600$ & 12 & 34\\
Q47-$\eta4$-$\zeta0$ &3.34E+47&1.E-04&1&1.14E+37&0.9997&38.7757&0.2983&3.65E+5&$(400)^2\times600$ & 12 & 39\\
Q47-$\eta3$-$\zeta0$ &3.34E+47&1.E-03&1&1.19E+37&0.9973&13.6911&0.3033 &3.59E+5&$(400)^2\times600$ & 12 & 35\\
\hline
Q45-$\eta4$-$\zeta1$ &3.34E+45&1.E-04&10&1.20E+35&0.9157&2.4881&0.0409 &2.66E+6&$(400)^2\times600$ & 12 & 50\\
Q46-$\eta4$-$\zeta1$ &3.34E+46&1.E-04&10&1.15E+36&0.9905&7.2607&0.1188 &9.16E+5&$(400)^2\times600$ & 12 & 45\\
Q47-$\eta4$-$\zeta1$ &3.34E+47&1.E-04&10&1.13E+37&0.9990&22.5645&0.2972 &3.66E+5&$(400)^2\times600$ & 12 & 38\\
\enddata
\tablenotetext{a}{Simulation resolution in terms of the jet radius, $r_j=1$ kpc.}
\tablenotetext{b}{Simulations end when the bow shock reaches either the top $z$-boundary or the side $x$ and $y$-boundaries.} 
\end{deluxetable*}

\subsection{Jet Setup}
\label{s2.3}

The jet is injected through the nozzle at $z=0$ with $\rho_j=\eta \rho_b$, $p_j=\zeta p_b$, and $v_j$ or $\Gamma_j = 1/\sqrt{1- (v_j/c)^2}$. Then, the jet power, $Q_j$, the amount of the kinetic plus internal energy (excluding the mass energy) injected into the background medium per unit time, is given as
\begin{equation}
Q_j=\pi r^{2}_jv_{j}\left(\Gamma^{2}_j\rho_jh_j-\Gamma_j\rho_jc^2\right).
\label{Qjet}
\end{equation}
This is the primary parameter that governs the morphological and dynamical evolution of jets through the interactions with the ICM, as mentioned in the introduction. The density and pressure ratios, $\eta$ and $\zeta$, are the secondary parameters, which may be combined into the momentum injection rate, or the jet thrust,
\begin{equation}
\dot{M}_j=\pi r^{2}_j\left(\Gamma_j^{2}\rho_j\frac{h_j}{c^2}v_j^{2}+p_j\right).
\label{Mjet}
\end{equation}
In this study, we specify the three parameters, $Q_j$, $\eta$, and $\zeta$; then, $v_j$ (and $\Gamma_j$) and $\dot{M}_j$ are determined.

Table \ref{tab:t1} shows the parameters of the jet models considered here. The first column lists the model name, which consists of three elements, the exponents of $Q_j$, $\eta$, and $\zeta$. The three fiducial models in the first group have the same $\eta=10^{-5}$ and $\zeta=1$, but different $Q_j$'s, Q45, Q46, and Q47. They are intended to represent the low-power, intermediate-power, and high-power jets of FR-II radio galaxies, respectively, and the Lorentz factor of jet flows ranges $\Gamma_j \approx 7.3 -71$. Those attached with H denote the high-resolution models. High-resolution simulations have been run only for the Q46 and Q47 models, owing to the much longer computational time required for the Q45 model. The models in the second group include the jets of higher densities with $\eta=10^{-4}-10^{-3}$, while those in the third group include over-pressured jets with $\zeta=10$. In our jet models, for a fixed value of $Q_j$, the higher density (larger $\eta$) or higher pressure (larger $\zeta$) of the jet leads to smaller $v_j$ or $\Gamma_j$, whereas larger $\eta$ or smaller $\zeta$ results in larger $\dot{M}_j$. Note that for very high power jets with large $\Gamma_j$, $\dot{M}_j\approx Q_j/c$. The adiabatic index of injected jet material is fixed by the ratio of $\eta/\zeta$; for $\eta/\zeta=10^{-5}$, the temperature is relativistic with $p_j/\rho_jc^2 = 0.764$ and $\gamma=1.43$, while for $\eta/\zeta=10^{-3}$, $\gamma\approx5/3$. We point that the temperature of shocked jet material in the hot spot is higher than that of injected material, and the adiabatic index there approaches $\gamma=4/3$ in some of our models.

For stable FR-II type jets, assuming a balance between the jet ram pressure and the background pressure, the advance speed of the jet head was estimated approximately as
\begin{equation}
v_{\rm head}^* \approx \frac{\sqrt{\eta_{r}}}{\sqrt{\eta_{r}}+1}v_j
\label{Vhead}
\end{equation}
\citep{marti1997}. Here, $\eta_{r} = (\rho_{j}h_{j}\Gamma^{2}_j)/(\rho_b h_b)$ is the relativistic density contrast. In general, $\eta_{r}> \eta$, but $\eta_{r}$ approaches $\eta$ for non-relativistic jet speeds and internal energies. We define the jet crossing time as
\begin{equation}
t_{\rm cross} = \frac{r_j}{v_{\rm head}^*},
\label{tcross}
\end{equation}
which can be used as a characteristic timescale for the jet evolution. The values of $v_{\rm head}^*$ and $t_{\rm cross}$ are given in the 8th and 9th columns of Table \ref{tab:t1}. Both $v_{\rm head}^*$ and $t_{\rm cross}$ are fixed mainly by $Q_j$; for instance, $t_{\rm cross}$ for Q45 is $6-7$ times longer than $t_{\rm cross}$ for Q47. Although $v_{\rm head}^*$ and $t_{\rm cross}$ depend also on $\eta$ and $\zeta$ (and $\dot{M}_j$), the dependence is weak. Below the results of jet simulations are described in terms of $t_{\rm cross}$.

The jet flow is directed upward mostly along the $z$-axis with $v_z \approx v_j$. However, to break the rotational symmetry, a slow, small-angle precession with period $\tau_{\rm prec}=10~t_{\rm{cross}}$ and angle $\theta_{\rm prec}=0.5^{\circ}$ is applied to the jet velocity. Also to ensure a smooth launching of jets, preventing possible developments of unphysical structures in the start-up, the jet velocity is modified with a window function as
\begin{equation}
v^*_{j} = v_{j}\left[1-(r/r_{j})^n\right]~~{\rm with}~~n=20,
\end{equation}
in the early stage of simulations. Here, $r=\sqrt{x^2+y^2}$ is the radial distance from the jet axis. The windowing is applied for one jet crossing time, $t=t_{\rm cross}$, and turned off afterward.

\subsection{Simulation Code}
\label{s2.4}

For the models listed in Table \ref{tab:t1}, simulations have been carried out using the newly developed 3D RHD code presented in Paper I. The version used for this study includes (1) a 5th-order accurate, finite-difference WENO scheme, WENO-Z \citep{borges2008}, (2) a 4th-order accurate time-integration method, the strong stability preserving Runge–Kutta (SSPRK) \citep{spiteri2003}, and (3) a 4th-order accurate averaging of fluxes along the transverse directions in smooth flow regions, which improves the performance in  multi-dimensional problems involving complex flows \citep{buchmuller2014}. In addition, to suppress the carbuncle instability, which often appears at the bow shock surface, the code incorporates a modification of eigenvalues for the acoustic modes; the local sound speed is limited to ${c}'_{s} = \min(\phi|{v}_{x}|,{c}_{s})$ in the calculation of eigenvalues \citep{fleischmann2020}. For the Q46 and Q47 models, $\phi=10$ is used, while $\phi=5$ is used for the Q45 models. For the CFL condition, ${\rm CFL}=0.8$ is used. The code implements Message Passing Interface (MPI) for parallel computing. The details of numerical schemes and performance tests can be found in Paper I.

Simulations have been performed in the boxes elongated along the $z$-direction, which consist of either $(400)^2\times600$ (for default models) or $(600)^2\times1200$ (for high-resolution models) uniform grid zones. The jet radius $r_j=1$ kpc occupies either $N_j=12$ grid zones (for default models) or 18 grid zones (for high-resolution models). Hence, the size of grid zones is $\Delta x=1/12$ kpc  or $\Delta x=1/18$ kpc, and the volume of the simulation domain is $(33.3)^2\times50$ kpc$^3$ or $(33.3)^2\times66.7$ kpc$^3$. The number of grid zones and the resolution in terms of the jet radius are listed in the 10th and 11th columns of Table \ref{tab:t1}.

It is known that the imposed condition at the bottom boundary of $z=0$ affects the properties of simulated jets \citep[e.g.,][]{donohoe2016}. The commonly used condition is either ``outflow'' or ``reflection''. With the outflow condition, some of the material along with the energy and momentum escapes from the simulation domain (see Figure \ref{f4} below), while it is conserved with the reflection condition. As a consequence, for instance, the jet morphology turns out to be more elongated with the outflow condition than with the reflection condition \citep[e.g.,][]{kossl1988}. We apply the continuous outflow condition to all the six faces of the simulation domain, including the $z=0$ plane except at the jet nozzle. Simulations stop when the bow shock reaches either the top $z$-boundary or the side $x$ and $y$-boundaries, and the end time of simulations, $t_{\rm end}$, is given in the last column of Table \ref{tab:t1}.

\begin{figure}[t] 
\centering
\vskip 0.1 cm
\includegraphics[width=0.55\linewidth]{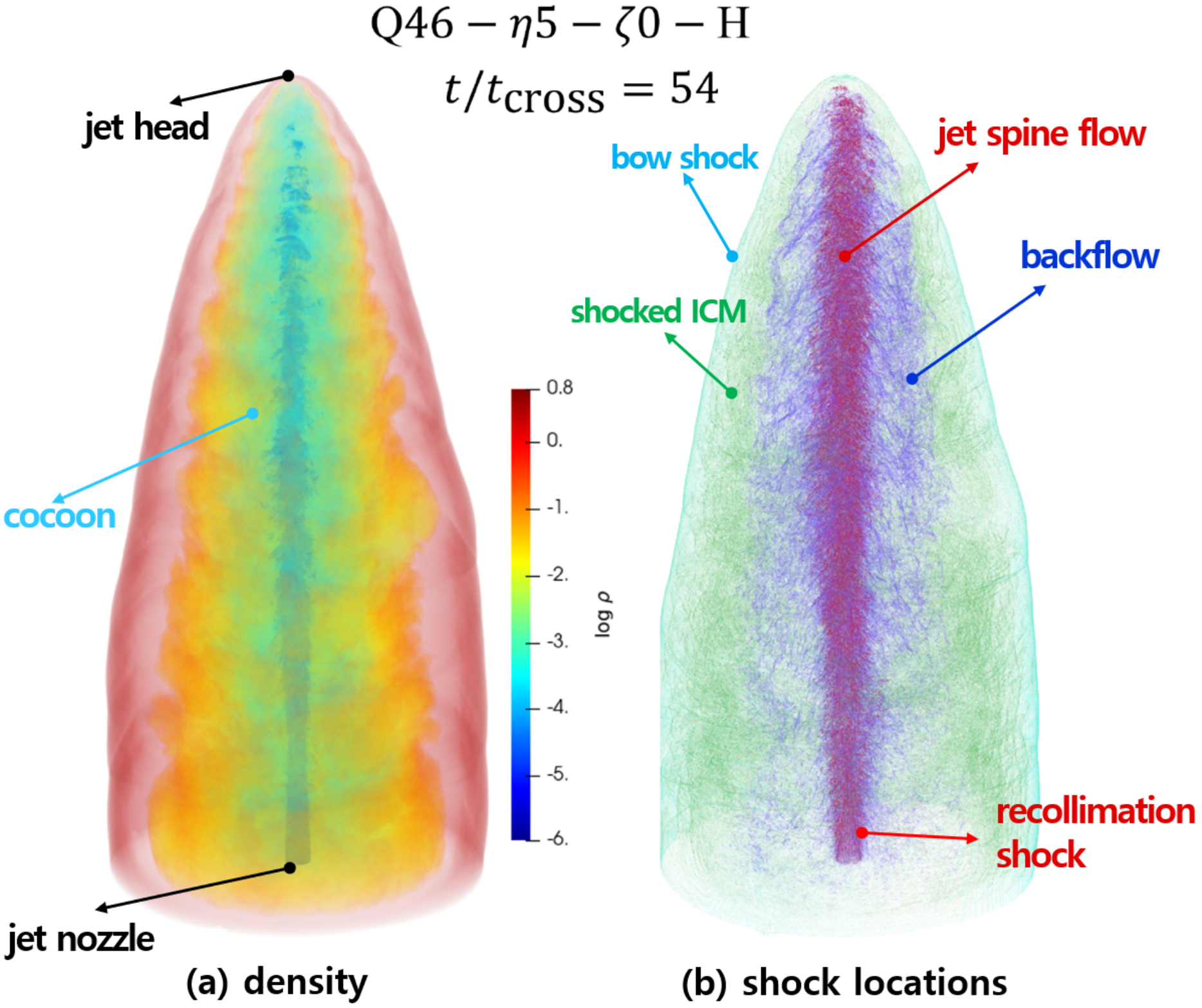}
\vskip -0.1 cm
\caption{(a) 3D volume-rendered image of $\log\rho$ in the Q46-$\eta5$-$\zeta0$-H model. Different regions are shown with different colors: the jet spine flow with bluish, the backflow with cyan to yellow, and the shocked ICM with reddish. (b) Spatial distribution of shocks in the same model. The locations of shock zones are color-coded: red in the jet spine flow, blue in the backflow, green in the shocked ICM, and cyan for the bow shock. While shocks with $M_{s}\ge 1.01$ are identified, in the jet spine flow and backflow, only those with $M_{s}\ge 1.1$ are shown for clarity. See Table \ref{tab:t1} for the model parameters.}
\label{f1}
\end{figure}

\begin{figure*}[t] 
\centering
\vskip 0.1 cm
\includegraphics[width=0.78\linewidth]{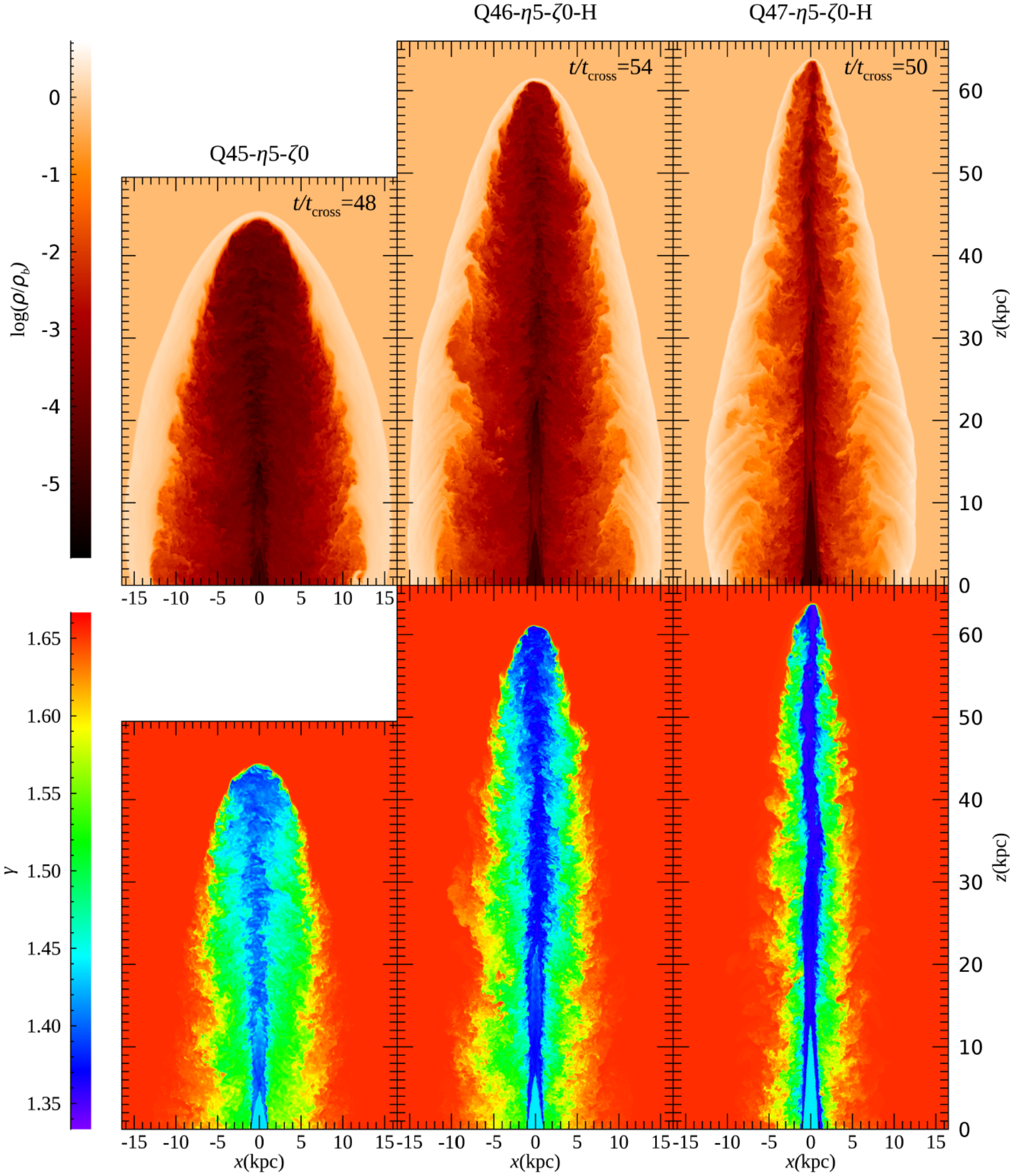}
\vskip -0.1 cm
\caption{2D slice images of $\log\rho$ (top panels) and the adiabatic index, $\gamma$ (bottom panels), through the plane of $y=0$ for the three fiducial models with different jet powers, Q45-$\eta5$-$\zeta0$, Q46-$\eta5$-$\zeta0$-H, and Q47-$\eta5$-$\zeta0$-H, at $t=t_{\rm end}$. See Table \ref{tab:t1} for the model parameters. In the top panels, the color changes from dark red for the light jet spine flow, red for the backflow, to whitish yellow for the dense shocked ICM. The background ICM is presented with peach yellow. In the bottom panels, the adiabatic index varies from $\gamma\approx4/3$ (purple-blue) around the jet head to $\gamma\approx5/3$ (red) in the background and shocked ICM. While the Q46 and Q47 models have $(600)^2\times1200$ grid zones, the Q45 model has $(400)^2\times600$ grid zones and hence its box has a different axial ratio.}
\label{f2}
\end{figure*}

\begin{figure*}[t] 
\centering
\vskip 0.1 cm
\hskip -0.2 cm
\includegraphics[width=0.9\linewidth]{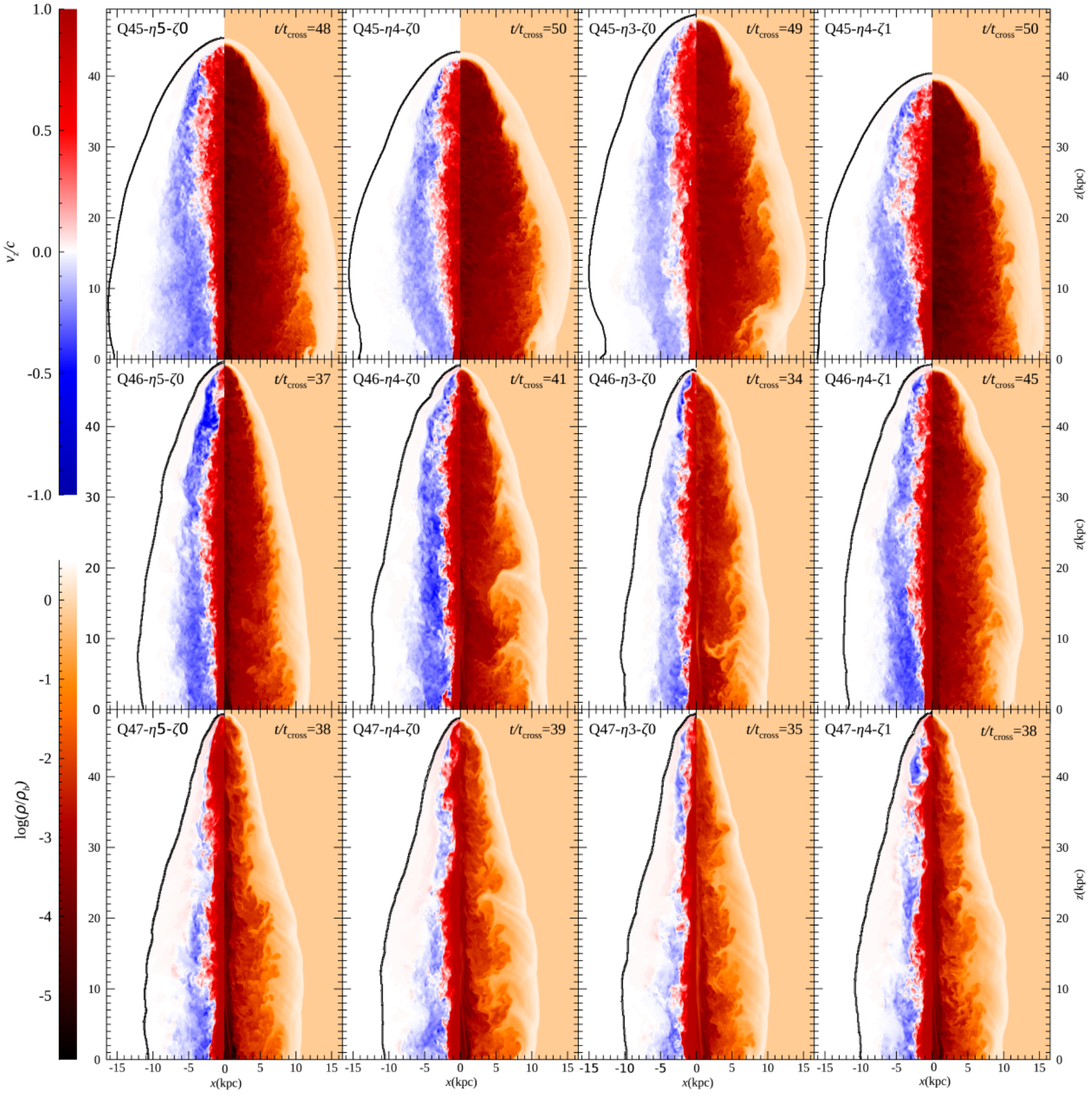}
\vskip -0.1 cm
\caption{2D slice images of $v_z$ on the left side ($x<0$) and $\log\rho$ on the right side ($x>0$) through the plane of $y=0$ in the Q45 (top panels), Q46 (middle panels), and Q47 (bottom panels) models with different $\eta$ and $\zeta$ at $t=t_{\rm end}$. See Table \ref{tab:t1} for the model parameters. On the left side, the upward-moving flow ($v_z>0$), which is mostly the jet spine flow, is shown in red, and the downward-moving flow ($v_z<0$), which is mostly the backflow, is shown in blue, whereas the shocked ICM ($v_z\sim0$) is shown in white. The bow shock is plotted with black dots to distinguish the shocked ICM from the background ICM. On the right side, $\log\rho$ is shown with the same color scheme as in Figure \ref{f2}.}
\label{f3}
\end{figure*}

\section{Jet Structures}
\label{s3}

\subsection{Morphology of Jets}
\label{s3.1}

Previous numerical studies have shown that the characteristic morphology of light, relativistic, FR-II-type jets may include the following features: (1) a terminal shock (or ``working surface'') at the head of the jet where the flow is halted and reversed, (2) a bow shock plowing through the denser background medium and representing the outer surface of the entire jet structures, (3) a board cocoon of the shocked jet material flowing backward, (4) the shock-heated background medium encompassing the cocoon, and (5) recollimation shocks formed in the jet spine flow \citep[e.g.,][]{english2016,li2018,perucho2019}. In reality, a stable distinct terminal shock may not appear, because the flows in the head become turbulent \citep[e.g.,][]{hardcastle2013,li2018}. Note that the cocoon filled with relativistic plasma is expected to be observed as radio lobe, so the two terms, cocoon and lobe, are often used interchangeably in describing jet structures.

In our simulated jets, we classify the structures bounded by the bow shock into three regions: (1) the highly under-dense, jet spine flow with $\rho < 0.1~\rho_b$ and $v_z > 0.1~c$, which is injected from the nozzle and keeps focused by recollimation shocks, (2) the low density, backflowing jet material with $\rho < 0.1~\rho_b$ and $v_z < 0.1~c$, which is halted and reversed at the jet head, and (3) the higher density, shocked ICM with $\rho > 0.1 \rho_b$ behind the bow shock. The left panel of Figure \ref{f1} illustrates those regions in one of our jet models, Q46-$\eta5$-$\zeta0$-H, moving from the inside to the outside, the jet spine flow, the backflow, the shocked ICM, and the bow shock, which is the outermost surface (see also the top panels of Figure \ref{f2}).

Figure \ref{f2} shows the two-dimensional (2D) slice images of $\log\rho$ and the adiabatic index, $\gamma$, for the three fiducial models, Q45-$\eta5$-$\zeta0$, Q46-$\eta5$-$\zeta0$-H, and Q47-$\eta5$-$\zeta0$-H. The density distributions in the top panels demonstrate how the morphology of jets depends on the jet power $Q_j$. Comparing the three fiducial models, one can see that in the models with higher $Q_j$ (and larger $\Gamma_j$), the jet advances faster in terms of $t_{\rm cross}$, resulting in more elongated structures. Note that the jet crossing time, $t_{\rm cross}$, is even shorter in the models with higher $Q_j$ (see Table \ref{tab:t1}). On the other hand, more extended cocoons of backflow, filled with shocks and turbulence, develop in the models with lower $Q_j$. The backflow mixes with the shocked ICM, as also can be seen in the distributions of $\gamma$ in the bottom panels. At the jet head, the adiabatic index is close to $\gamma=4/3$ (blue); in the backflow, it increases due to the turbulent mixing as well as the adiabatic expansion, and smoothly merges to the value of the shocked ICM, $\gamma=5/3$ (red).

\begin{figure}[t] 
\centering
\vskip 0.1 cm
\includegraphics[width=0.45\linewidth]{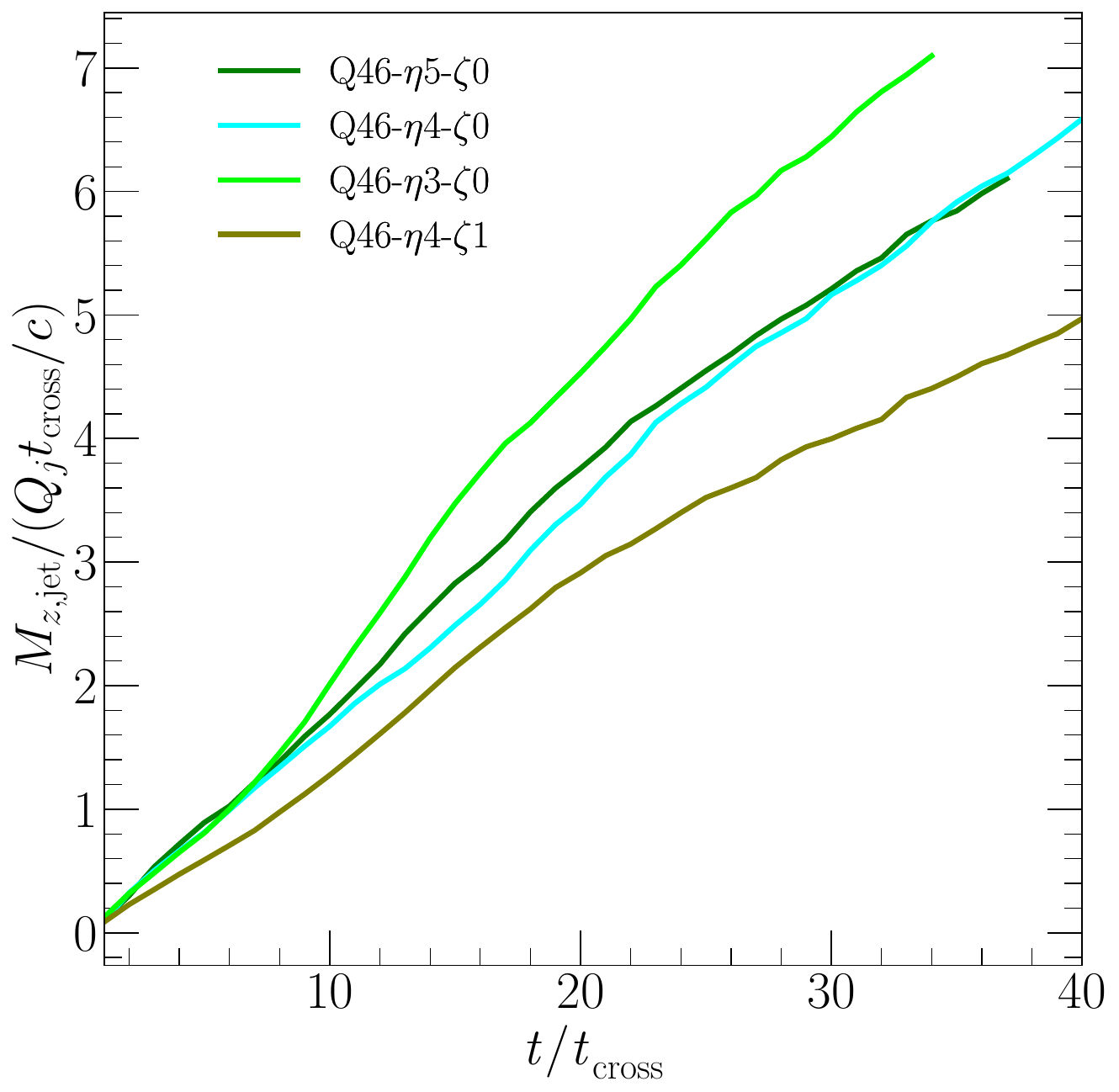}
\vskip -0.1 cm
\caption{Accumulated $z$-momentum contained in the jet spine flow, $M_{z,\rm{jet}}$, as a function of time in the Q46 models with different values of $\eta$ and $\zeta$. The three models, Q46-$\eta4$-$\zeta0$, Q46-$\eta5$-$\zeta0$, and Q46-$\eta4$-$\zeta1$, have similar momentum injection rates, $\dot{M}_j\approx 1.13-1.19\times 10^{36}~{\rm dyne}$, while Q46-$\eta3$-$\zeta0$ has a slight higher rate, $\dot{M}_j\approx 1.37\times 10^{36}~{\rm dyne}$. }
\label{f4}
\end{figure}

In the right half ($x>0$) of each panel in Figure \ref{f3}, $\log\rho$ is plotted for jets with different $\eta$ and $\zeta$. Overall, while the influence of $\eta$ and $\zeta$ on the jet morphology is less significant than that of $Q_j$, as already known from previous studies (see the introduction), the figure shows that those secondary parameters affect some properties of the jets, such as the advance speed, in our simulations. In the high-power Q47 models, the jet advance speed is not very sensitive to $\eta$ and $\zeta$, while the jet of Q47-$\eta3$-$\zeta0$ propagates a bit faster where the momentum injection rate, $\dot{M}_j$, is slightly larger (see Table \ref{tab:t1}). In the lower power models (Q46 and Q45), the jet expansion rate differs somewhat for different $\eta$ and $\zeta$. For instance, among the four Q46 models, the jet of Q46-$\eta3$-$\zeta0$ with the largest $\dot{M}_j$ advances the fastest. Although the other three Q46 jets have similar $\dot{M}_j$, they propagate with somewhat different speeds. We find that the amount of the $z$-momentum contained in the jet spine flow, $M_{z,\rm{jet}}$, which pushes the jet head forward, turns out to be different. Figure \ref{f4} shows the accumulated $M_{z,\rm{jet}}$ as a function of $t/t_{\rm cross}$ for the four Q46 models. Even though $\dot{M}_j$ is similar among the three models, $M_{z,\rm{jet}}$ may develop differently owing to different dynamical evolutions. For instance, in the Q46-$\eta4$-$\zeta1$ model where the jet pressure is higher, the amount of the material that escapes from the simulation domain through the $z=0$ boundary is less than in the Q46-$\eta5$-$\zeta0$ and Q46-$\eta5$-$\zeta0$ models, because the jet expands more laterally (see the middle panels of Figure \ref{f3}). The $z$-momentum in the whole jet structure as well as $M_{z,\rm{jet}}$ in the jet spine are smaller in the Q46-$\eta4$-$\zeta1$ model, since less material with negative $z$-momentum escapes from the system. Hence, the difference in the advance speed should be attributed at least partly to the outflow condition we impose at the $z=0$ boundary. A similar trend is found for the Q45 models.

In the left half ($x<0$) of each panel in Figure \ref{f3}, the vertical velocity, $v_z$, is plotted. The upward-moving jet spine flow is shown in red, while the downward-moving backflow is shown in blue. The shocked ICM surrounding the cocoon has relatively small vertical velocities ($v_z\sim 0$, white), while it expands mostly in the lateral direction behind the bow shock. The interface between the jet spine flow and the backflow and also the interface between the backflow and the shocked ICM become turbulent via the Kelvin-Helmholtz (KH) instability due to strong velocity shear. Moreover, the working surface is not very apparent owing to turbulent flows and the small precession in the injected flow at the jet nozzle.

A notable point in the $\log\rho$ images of Figures \ref{f2} and \ref{f3} is that the bow shock surface includes kink-like structures in the Q46 and Q47 jets, while it is relatively smooth in the low-power Q45 jets. In addition, structures resembling herringbone patterns appear in the density of the shocked ICM behind the bow shock for the Q46 and Q47 jets. Such structures were observed in previous simulations of high-power jets using high-accurate codes \citep[e.g., see][]{perucho2019}. Below, we will see similar patterns also emerge in the vorticity distribution. We expect that the development of these structures would be the consequence of the interaction between the bow shock and the turbulent flows in the cocoon.

\begin{figure*}[t] 
\centering
\vskip 0.1 cm
\hskip -1 cm
\includegraphics[width=1.04\linewidth]{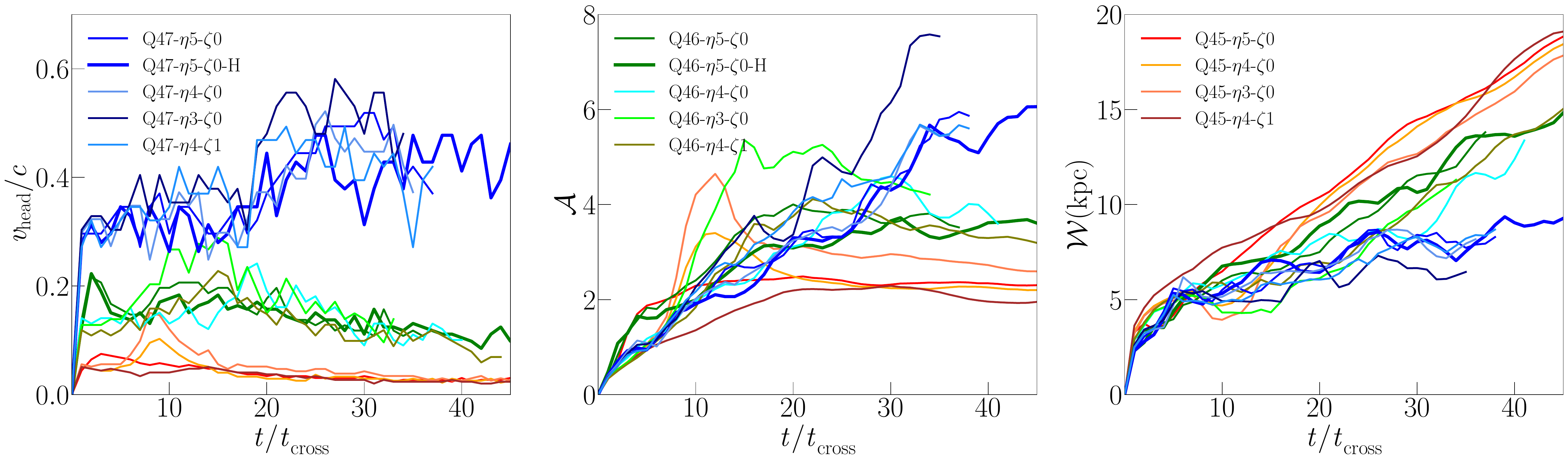}
\vskip -0.1 cm
\caption{Time variations of the jet-head advance speed, $v_{\rm head}$ (left panel), the lobe axial ratio, $\mathcal{A}=\mathfrak{L}/\mathcal{W}$ (middle panel), and the lateral width of the lobe, $\mathcal{W}$ (right panel), for all the models listed in Table \ref{tab:t1}. Here, $\mathfrak{L}$ is the length of the lobe. $\mathfrak{L}$ and $\mathcal{W}$ are obtained using the simulation data. The thick lines are for high-resolution models. The models with the same $Q_j$ are shown in the same hue of colors, reddish for Q45, greenish for Q46, and bluish for Q47.}
\label{f5}
\end{figure*}

As the jet propagates into the background medium, while the whole jet-induced structure expands, the head where the jet spine flow stops advances. After the initial adjustment, the advance speed of the jet head is expected to approach the analytic estimate in Equation (\ref{Vhead}), roughly $v_{\rm head}^*\approx0.05c$, $0.1c$, and $0.3c$ for the Q45, Q46, and Q47 models, respectively (see Table \ref{tab:t1}). The left panel of Figure \ref{f5} shows the time evolution of the advance speed, $v_{\rm head}$, which is determined with the actual position of the jet head in simulations, for all the models considered here. The values of $v_{\rm head}$ fluctuate around $v_{\rm head}^*$, but after $t/t_{\rm cross} \sim 10-20$, they tends to approach asymptotic values. We find that the asymptotic values are roughly $v_{\rm head}\approx0.025c$, $0.1c$, and $0.45c$ for the Q45, Q46, and Q47 models, respectively. That is, the numerically estimated asymptotic values are somewhat larger than $v_{\rm head}^*$ in the high-power Q47 models, while they are smaller than $v_{\rm head}^*$ in the low-power Q45 models. In the Q46 models, the asymptotic values are close to $v_{\rm head}^*$.

The shape of the lobe (cocoon) may be quantified with the axial ratio, $\mathcal{A}\equiv \mathfrak{L}/\mathcal{W}$, where $\mathfrak{L}$ is the vertical length of the cocoon and $\mathcal{W}$ is the lateral width at its midpoint \citep{hardcastle2013}. The middle panel of Figure \ref{f5} shows $\mathcal{A}$, which is measured from the simulation results. The axial ratio undergoes variations due to the competition between the longitudinal advancement and the lateral expansion. Overall, $\mathcal{A}$ is larger if the jet is more powerful. In the high-power Q47 models, the shape of the cocoon is highly elongated with $\mathcal{A} \sim 6$ or even larger, which still increases at $t_{\rm end}$ in our simulations.

In the Q45 and Q46 models, on the other hand, $\mathcal{A}$ on average increases up to $t/t_{\rm cross} \sim 20$ and then approaches asymptotic values. In these models, while the jet advances slowly, the over-pressured cocoon expands laterally, resulting in smaller asymptotic axial ratios; roughly $\mathcal{A} \sim2$ and $\sim4$ for the Q45 and Q46 models, respectively. As noted above, $v_{\rm head}$, the increment speed of $\mathfrak{L}$, is about four times larger in the Q46 models than in the Q45 models. The lateral expansion speed, $v_{\rm lateral}$, could be estimated from $\mathcal{W}$ shown in the right panel of Figure \ref{f5}, where the slope gives $v_{\rm lateral}\times t_{\rm cross}$. We find that $v_{\rm lateral}$ is about twice larger in the Q46 models. Since $v_{\rm head}/v_{\rm lateral}$ is about twice larger, we get $\mathcal{A}$ that is about twice larger in the Q46 models than in the Q45 models. In general, both $v_{\rm head}$ and $v_{\rm lateral}$ tend to increase with the jet power.

Comparing the thin and thick green lines for Q46-$\eta5$-$\zeta0$ and the thin and thick blue lines for Q47-$\eta5$-$\zeta0$ in Figure \ref{f5}, we see that the overall morphology of the jets is well converged in the default and high-resolution simulations (see also Figures \ref{f2} and \ref{f3}). Yet, we expect that finer structures and more flow motions develop in smaller scales in higher resolutions (see the next section).

As noted above, the jet is injected to the uniform background in our simulations, assuming that the jet propagation is confined within the cluster core region. In the case of giant radio galaxies that expand out to several 100~kpc into stratified halos, the jet head is expected to be decelerated due to the lateral expansion; then, the evolutionary trend of $\mathcal{A}$ may differ somewhat from that shown in Figure \ref{f5}.

In short, the morphology of the jet including the shape of the cocoon is primarily governed by the jet power, $Q_j$, and less dependent on $\eta$ and $\zeta$. Hence, we present mainly the fiducial models with different $Q_j$ in describing jet flow dynamics in the next section.  The other models will be used to examine the parameter dependence of the problem.

\begin{figure*}[t] 
\centering
\vskip 0.1 cm
\includegraphics[width=0.85\linewidth]{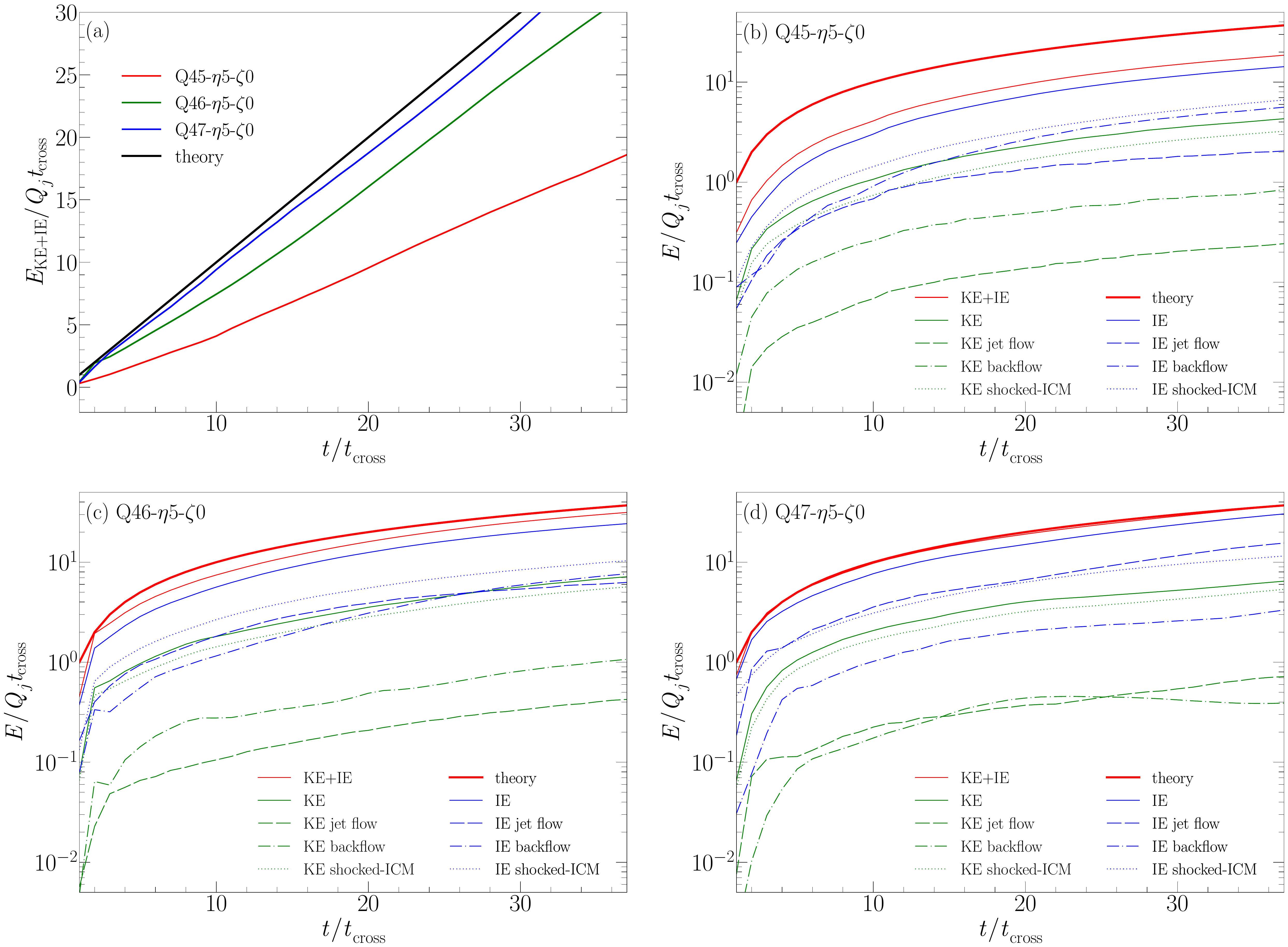}
\vskip -0.1 cm
\caption{Energies in the jet. (a) The kinetic + internal energy contained in the region encompassed by the bow shock surface, $E_{\rm KE+IE}$, is shown as a function time for the fiducial models with different $Q_j$. For comparison, the black line shows the kinetic + internal energy jet-injected into the simulation domain (labeled as theory). (b)-(d) The kinetic energy (KE, in blue) and internal energy (IE, in green) in the different regions, the jet spine flow, backflow, and shocked ICM, are shown as a function time for the fiducial models. In panel (d), the thin red line (KE+IE) almost coincides with the thick red line (theory). See Table \ref{tab:t1} for the model parameters.}
\label{f6}
\end{figure*}

\subsection{Energetic of Jets}
\label{s3.2}

Before we describe jet flow dynamics, we examine the kinetic and internal energies contained in the different regions of the jet. We point that while the RHDs is formulated based on the energy-momentum conservation, the strict decomposition of the energy into the kinetic and internal energies is not plausible \citep[e.g.,][]{landau1959}. Yet, following previous works \citep[e.g.,][]{english2016}, we divide the energy in Equation (\ref{Econs}) as follows:
\begin{equation}
E = \Gamma(\Gamma-1)\rho c^2 + \left[\Gamma^2(h-c^2)\rho-p\right] + \Gamma\rho c^2,
\label{Edecomp}
\end{equation}
where the three terms in the right hand side could be regarded as the kinetic, internal, and mass energies, respectively. In terms of the adiabatic index in Equation (\ref{hIDEOS}), the internal energy can be written as $\gamma\Gamma^2 p/(\gamma-1)-p$. In the non-relativistic limit, the kinetic and internal energies reduce to the usual forms, $(1/2)\rho v^2$ and $p/(\gamma-1)$, respectively.

Panel (a) of Figure \ref{f6} shows the kinetic plus internal energy, $E_{\rm KE+IE}$ (excluding the mass energy), contained in the region encompassed by the bow shock surface, relative to the accumulated energy injected through the jet nozzle (labeled as theory), for three fiducial models. Note that the initial internal energy of the background material should be counted for the exact estimation of the expected energy, but it is small in our jet models, $\sim5~\%$, $\sim1~\%$, and $\sim0.2~\%$ of $E_{\rm KE+IE}$ around $t_{\rm end}$ for the Q45, Q46, and Q47 models, respectively. We find that the energy inside the bow shock surface is smaller than the injected energy, because a part of the energy escapes through the outflow boundary at $z=0$ \citep[see, also][]{english2016}. The fraction of the escaped energy is greater for lower $Q_j$, since broader, more turbulent cocoons develop.

Panels (b)-(d) of Figure \ref{f6} show the kinetic and internal energies in the different regions of the jet for the three fiducial models. 
The energy partitioning differs for different models. Roughly, inside the bow shock surface, the kinetic energy is several times smaller than the internal energy. The kinetic plus internal energy in the cocoon (including both the jet spine flow and backflow) is somewhat smaller than that in the shocked ICM. The kinetic energy in the cocoon, which is the manifested quantity in the flow dynamics described below, is always a small fraction of the jet energy; it is estimated to be $\sim 5~\%$, $\sim 3~\%$, and $\sim 2~\%$ of $E_{\rm KE+IE}$ inside the bow shock surface around $t_{\rm end}$ for the Q45, Q46, and Q47 models, respectively.

\section{Jet Flow Dynamics}
\label{s4}

\subsection{Shock Analysis}
\label{s4.1}

The jet-induced structure contains two types of shock surfaces in our simulations: (1) distinct, connected surfaces such as the bow shock and recollimation shocks, and (2) less prominent, somewhat disordered shock surfaces in turbulent flows such as the jet spine flow, backflow, and the shocked ICM (see Figure \ref{f8}).

Each shock surface is composed of many ``shock zones'' (numerical grid elements), which are identified in a post-processing step by applying a widely used algorithm \citep[e.g.,][]{ryu2003}. Along each coordinate axis, grid zones are tagged as ``shocked'', if they satisfy the following conditions: (1) $\mbox{\boldmath$\nabla$}\cdot\mbox{\boldmath$v$} < 0$, i.e., the locally converging flow, and (2) $\Delta p/p > \epsilon_{p}$, i.e., the pressure jump in the adjacent zones larger than $\epsilon_{p}p$. Here, $\epsilon_{p}$ is a free parameter that should depend on the minimum Mach of the shocks to be identified. In our simulations, each shock transition spreads typically over two to three ``shocked'' zones, so the zone with the minimum value of $\mbox{\boldmath$\nabla$}\cdot\mbox{\boldmath$v$}$ is identified as a ``shock zone''. The preshock and postshock states are then estimated across the shock transition. With the density and pressure in the preshock and postshock regions, the Mach number along each coordinate axis is calculated (see below); the Mach number of the shock zone is obtained as $M_s = \max(M_{s,x},M_{s,y},M_{s,z})$.

We here find shocks with $M_s\geq1.01$, expecting that shocks with $M_s < 1.01$ would be dynamically unimportant (see the next subsection). For non-relativistic shocks of $M_s=1.01$, $\Delta p/p\approx0.025$ across the shock transition. So we set $\epsilon_{p}=0.005$, since the shock transition spreads over grid zones.

\begin{figure*}[t] 
\centering
\vskip 0.1 cm
\includegraphics[width=0.85\linewidth]{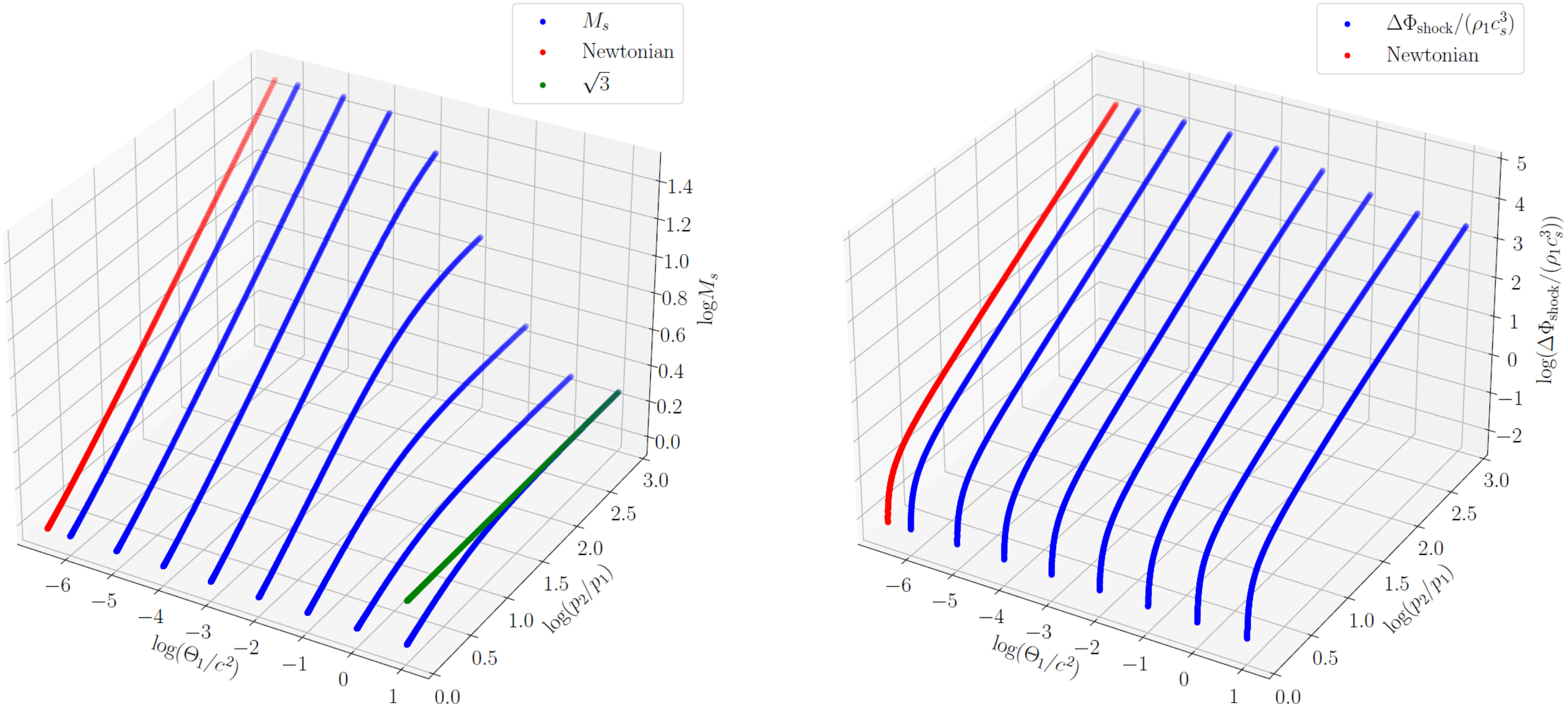}
\vskip -0.1 cm
\caption{$M_{s}$ (left panel) and $\Delta\Phi_{\rm shock}/(\rho_1c_{s,1}^3)$ (right panel) as a function of $p_2/p_1$ and $\Theta_1$ for RHD shocks in the shock-rest frame. The blue lines draw $M_s$ and $\Delta\Phi_{\rm shock}/(\rho_1c_{s,1}^3)$ for several representative values of $\Theta_1$. The red lines plot $M_s$ and $\Delta\Phi_{\rm shock}/(\rho_1c_{s,1}^3)$ for non-relativistic shocks with $\Theta_1 \ll 1$, and the green line in the left panel shows $M_{s}=\sqrt{3}$, the maximum $M_{s}$ for fully relativistic shocks with $\Theta_1\gg 1$.
}
\label{f7}
\end{figure*}

In Newtonian HDs, $M_{s}$ can be estimated with the pressure ratio across the shock, $p_2/p_1$, using $p_2/p_1= (2\gamma M_{s}^2 -\gamma + 1)/(\gamma+1)$, where the adiabatic index $\gamma=5/3$ for thermally non-relativistic, monatomic gas. Hereafter, the subscripts 1 and 2 denote the preshock and postshock states, respectively. In RHDs, however, $M_{s}$ cannot be determined by $p_2/p_1$ alone. In addition, $M_{s}$ is a quantity that depends on the frame where it is obtained.

From the density, momentum, and energy conservations across the shock, the shock speed can be expressed as
\begin{equation}
\frac{v_{s}}{c} = \sqrt{\frac{1-h_{1}^{2}/h_{2}^{2}}{1-\rho_{1}^{2}/\rho_{2}^{2}}},
\label{vshock}
\end{equation}
in the shock-rest frame. On the other hand, for instance, if the shock moves with $v_t$ along the direction transverse to the shock normal, the shock speed is modified to $v_{s}/{\Gamma_t}$, where $\Gamma_t=1/\sqrt{1-(v_t/c)^2}$. It is technically non-trivial to find the shock-rest frames for all the shock zones in our simulations, and hence it is practically difficult to estimate the velocity with which each shock zone moves in the computational frame. Thus, we calculate the Mach numbers of shock zones in the shock-rest frame, rather than in the computational frame. This would introduce uncertainties in the characteristic properties of identified shocks especially in the jet spine flow, as some of those have $v_s$ close to $c$. Such effects should not be substantial for shocks in other parts of the jet (see Figure \ref{f9}).

For identified shock zones, the estimation of $v_{s}$ with Equation (\ref{vshock}) using numerical values of $h_1/h_2$ and $\rho_1/\rho_2$ would not be robust, since those two ratios are close to unity and not very sensitive to $M_{s}$, especially at weak shocks. Instead, we find that it is more reliable to use the ratio $p_2/p_1$ in estimating $M_{s}$. Again, from the conservations across the shock, we can get
\begin{equation}
\frac{h_{1}}{c^2}\left(\frac{h_{2}^{2}}{h_{1}^{2}}-1\right)\frac{\rho_{1}}{\rho_{2}} = \left(\frac{p_{2}}{\rho_{2}c^2} - \frac{p_{1}}{\rho_{1}c^2}\frac{\rho_{1}}{\rho_{2}}\right)\left(1+\frac{h_{2}}{h_{1}}\frac{\rho_{1}}{\rho_{2}}\right).
\label{shockeq}
\end{equation}
Then, for given values of $p_2/p_1$ and $\Theta_1\equiv p_1/\rho_1c^2$, the ratios, $h_1/h_2$ and $\rho_1/\rho_2$, can be obtained using Equations (\ref{hRCEOS}) and (\ref{shockeq}), and the shock speed can be calculated from Equation (\ref{vshock}). With the RC EOS, the sound speed of the preshock gas is given as
\begin{equation}
\frac{c_{s,1}}{c}=\sqrt{\frac{\Theta_{1}(3\Theta_{1}+2)(18\Theta_{1}^{2}+24\Theta_{1}+5)}{3(6\Theta_{1}^{2}+4\Theta_{1}+1)(9\Theta_{1}^{2}+12\Theta_{1}+2)}}.
\label{csound}
\end{equation}
Hence, $M_{s} \equiv v_{s}/c_{s,1}$ can be calculated for given values of $p_2/p_1$ and $\Theta_1$. In practice, we have built a numerical table for $M_{s}$ as a function of $p_2/p_1$ and $\Theta_1$. The left panel of Figure \ref{f7} shows $M_s$ in the 2D parameter space of $(p_2/p_1,\Theta_1)$. We use this table for the estimation of $M_s$ with the numerical values of $p_2/p_1$ and $\Theta_1$ obtained for shock zones in simulated jets.

We then estimate the kinetic energy dissipation rate at the shock, $\Delta\Phi_{\rm shock}$, from the difference between the entering and leaving kinetic energy fluxes across the shock. It can be written in the shock-rest frame as
\begin{equation}
\begin{aligned}
\Delta\Phi_{\rm shock} \equiv \Gamma_{1}(\Gamma_{1}-1)\rho_{1}c^{2}v_{1} - \Gamma_{2}(\Gamma_{2}-1)\rho_{2}c^{2}v_{2}\\
= \frac{\rho_1c^{2}v_s}{1-(v_s/c)^2}\left(1-\frac{h_1}{h_2}\right),~~~~~~~~~~~~~~~~~
\label{keflux}
\end{aligned}
\end{equation}
where $v_1=v_s$, and $\Gamma_{1}$ and $\Gamma_{2}$ are the Lorentz factors of the preshock and postshock flow speeds, $v_1$ and $v_2$, respectively. It reduces to $(1/2)\rho_1v_s(v_s^2-v_{2}^2)$ in the non-relativistic limit. Below, we use $\Delta\Phi_{\rm shock}$ to estimate the total amount of the jet energy dissipated at shocks in the jet-induced flows. This quantity is also frame-dependent. For instance, if the shock moves with a transverse velocity of $v_t$, it is given as $\Delta\Phi_{\rm shock}\Gamma_t$. We here employ $\Delta\Phi_{\rm shock}$ in the shock-rest frame, rather than the one in the computational frame; hence our estimation of the shock dissipation should be considered only approximate.

As in the case of $M_s$, we have built a numerical table for $\Delta\Phi_{\rm shock}/(\rho_1c_{s,1}^3)$ as a function of $p_2/p_1$ and $\Theta_1$, which is plotted in the right panel of Figure \ref{f7}. We use this table for the estimation of $\Delta\Phi_{\rm shock}$ with the numerical values of $p_2/p_1$,$\Theta_1$, and $\rho_1$ obtained for shock zones in simulated jets.

\subsection{Properties of shocks}
\label{s4.2}

\begin{figure*}[t] 
\centering
\vskip 0.1 cm
\includegraphics[width=1\linewidth]{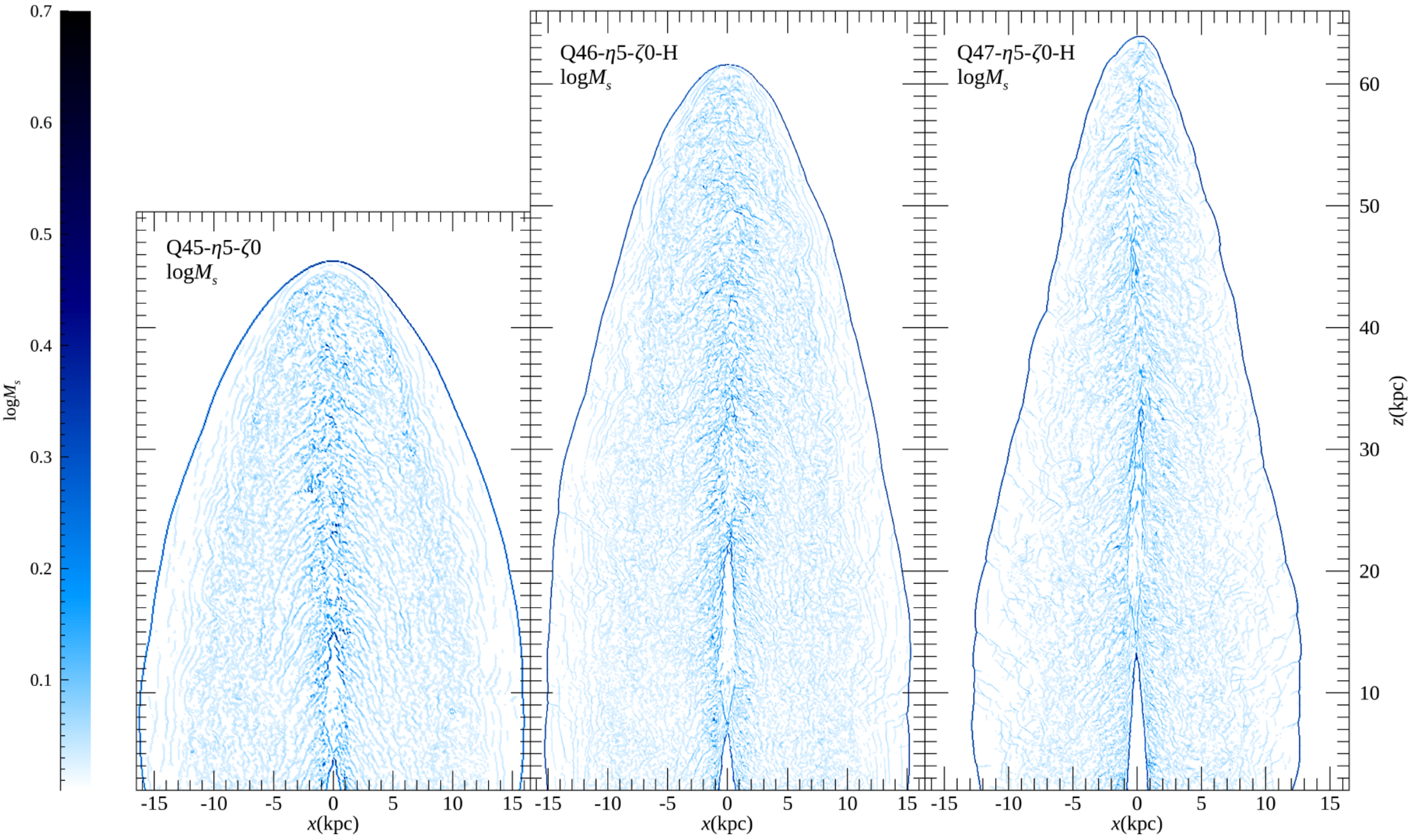}
\vskip -0.1 cm
\caption{2D slice distributions of shocks with $M_s\geq1.01$ through the plane of $y=0$ for the three fiducial models, Q45-$\eta5$-$\zeta0$, Q46-$\eta5$-$\zeta0$-H, and Q47-$\eta5$-$\zeta0$-H, at $t=t_{\rm end}$. See Figure \ref{f2} for the description of the axial ratio of the simulation domain.}
\label{f8}
\end{figure*}

\begin{figure*}[t] 
\centering
\vskip 0.1 cm
\includegraphics[width=1.0\linewidth]{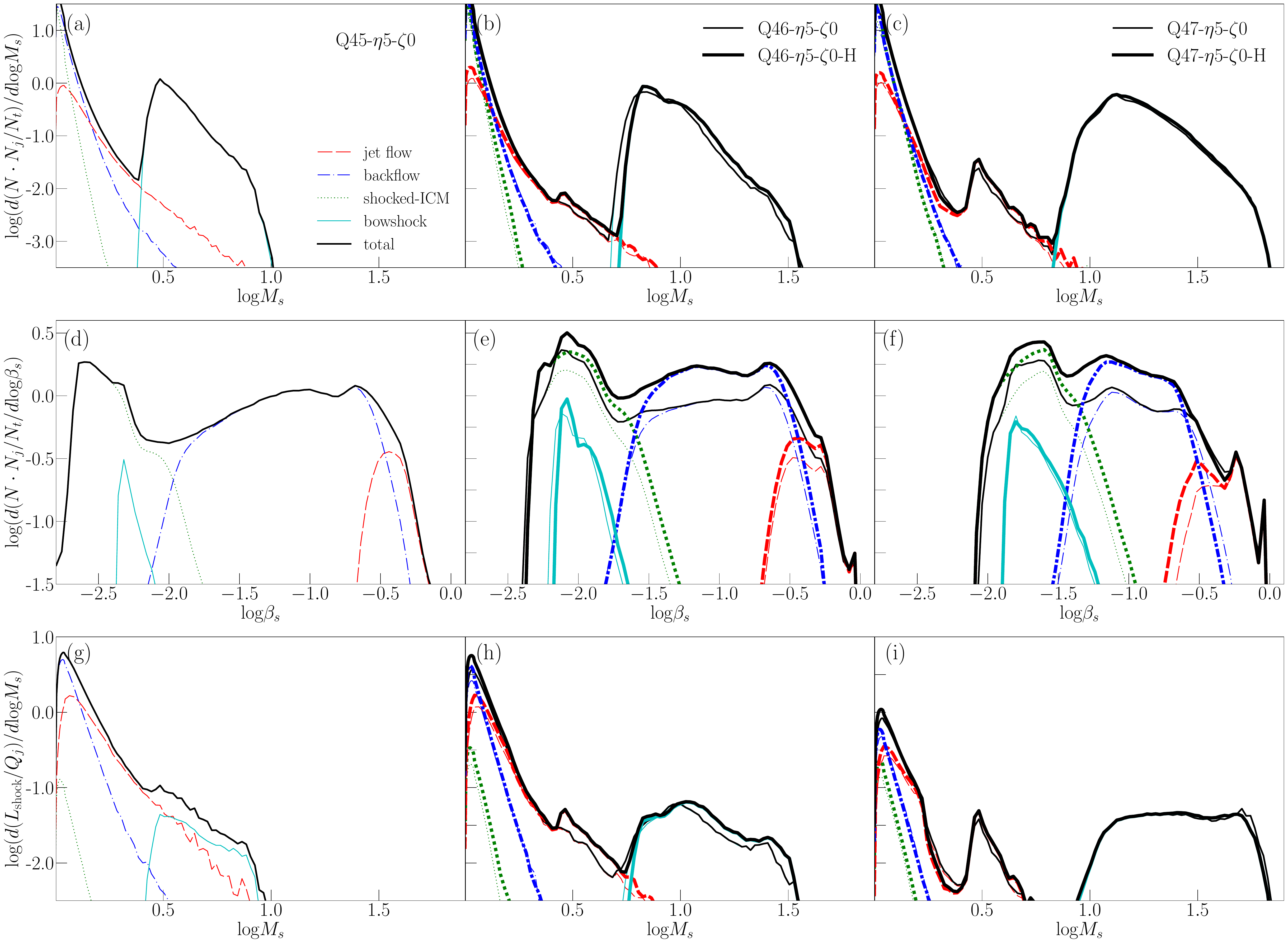}
\vskip -0.1 cm
\caption{PDFs of the shock Mach number, $M_s$ (top panels), PDFs of the shock speed, $\beta_{s}= v_{s}/c$ (middle panels), and the energy dissipation rate at shocks, $L_{\rm shock}(M_s)$, in Equation (\ref{lke}) (bottom panels) for the shock zones in the jet spine flow (red dashed lines), backflow (blue dot-dashed lines), and shocked ICM (green dotted lines), and for the bow shock surface (cyan solid lines). The black solid lines plot the quantities for all the shocks identified in the simulation domain. The shock zones with $M_s\geq1.01$ are included, and the quantities shown are averaged over $t/t_{\rm cross}=[20,30]$. All the five fiducial models in Table \ref{tab:t1} are shown. The thick lines in the middle and right panels are for high-resolution models. Here, $N_t$ is the total number of grid zones in the volume encompassed by the bow shock surface, and $N_j=r_j/\Delta x$ is the number of zones occupied by the jet radius. The shock dissipation is normalized by the jet energy, $Q_j$.}
\label{f9}
\end{figure*}

Panel (b) of Figure \ref{f1} illustrates the 3D spatial distribution of shock zones formed in the different regions of the Q46-$\eta5$-$\zeta0$-H jet. Figure \ref{f8} shows the 2D distributions of shock zones for three fiducial models. The bow shock surface surrounds the entire jet-induced structure and separates the shocked ICM from the background medium. While the so-called recollimation shocks appear along the jet spine, the first one is the most distinctive and its surface has a conical shape that stretches further upward from the jet nozzle in the models with higher $Q_j$. Although the surfaces of the bow shock and the first recollimation shock are composed of many shock zones with different $M_s$, we find that those shocks can be characterized with typical values of $M_s$ (see below). On the other hand, the surfaces of shocks formed in turbulent flows such as the backflow and shocked ICM are more chaotic and less distinct. While there are many of them, the sizes of connected shock surfaces are much smaller than that of the bow shock surface. We note that the termination shock does not clearly appear in our model jets due to the turbulence in the head and backflow. Below, we quantify the properties of shock zones to examine the characteristics of the shocks formed in the jet-induced flows.

Panels (a)-(c) of Figure \ref{f9} plot the probability distribution functions (PDFs) of the shock Mach number, $N(M_s)N_j/N_t$, for shock zones in the different regions of the jet for five fiducial models. Here, $N(M_s)$ is the number of shock zones with $M_s$ in the range of [$\log M_s$,$\log M_s+ d\log M_s$]. $N_{t}$ is the total number of grid zones in the volume encompassed by the bow shock surface, while $N_j=r_j/\Delta x$ is the number of grids across the jet radius. Note that $N(M_{s})$ is proportional to the area of shock surfaces, while $N_j$ and $N_t$ are proportional to the jet radius and the volume of the jet-induced structure; hence, $N(M_s)N_j/N_t$ is effectively a dimensionless quantity.

The Mach number is the highest for the bow shock (cyan) and the next highest for shocks in the jet spine flow (red), and it is relatively low for shocks in the backflow (blue) and shocked ICM (green). But this is not necessarily the order of the shock speed (see below), since $v_s=M_s c_{s,1}$ depends on the preshock sound speed, $c_{s,1}$, as well. While $c_{s,1} \ll c$ in the background ICM, it is much larger in the jet spine flow and backflow; in particular, $c_{s,1}$ is close to $c/\sqrt{3}$ in the regions where the adiabatic index is close to $\gamma=4/3$ (see Figure \ref{f2}).

Panels (a)-(c) of Figure \ref{f9} manifest two distinct populations of shocks: (1) the population for peaked PDFs with characteristic $M_{s}$'s, which consists of shock zones of the bow shock and recollimation shock surfaces, and (2) the population for power-law-like PDFs, which is associated with turbulent flows \citep[e.g.,][]{park2019}. The PDF for shock zones associated with the bow shock, peaks at $M_{s} \simeq 3.0$, 6.5, and 12.6 in the Q45, Q46, and Q47 models, respectively; the strength of the bow shock increases with increasing $Q_j$, as expected. The characteristic Mach number of shock zones associated with the first recollimation shock is $M_{s} \simeq 2.9$ and 3.0 in the Q46 and Q47 models. In the case of Q45-$\eta5$-$\zeta0$, the peak due to the recollimation shock is not noticeable, as it is expected to occur at smaller $M_s$. Again, the strength of the first recollimation shock increases with $Q_j$, which can be also guessed with the adiabatic index in the bottom panels of Figure \ref{f2}. On the other hand, shocks in turbulent parts of the jet-induced structure are relatively weak with $M_s\lesssim$ a few. Shock zones associated with these disordered shocks follow fairly steep power-law-like distributions, $N(M_{s})\propto M_{s} ^{-q}$; the slope ranges $q\sim 5-10$ for shocks of low $M_s$ in the jet spine flow, $q\sim 12 - 13$ in the backflow, and $q\sim 15 - 20$ in the shocked ICM. The averaged $M_s$ of these shocks is less dependent on $Q_j$.

The integration of the PDF for all the shocks (black lines) gives the total number of shock zones with $M_s\geq1.01$, $N_{\rm shock}$, and the ratio, $N_t/(N_{\rm shock}N_j)$, provides a measure for the mean distance between shock surfaces (in units of $r_j$) over the whole jet-induced structure. We point that $N_{\rm shock}$ and $N_t$ increase with time, but their ratio does not change much, once the jet-induced structures have more-or-less fully developed, for instance, at $t\gtrsim20~t_{\rm cross}$ (see Figure \ref{f5}). The mean distance measured at $t=30~t_{\rm cross}$ increases with $Q_j$ as $N_t/(N_{\rm shock}N_j)=0.43$, 0.46, and 0.57 for Q45-$\eta5$-$\zeta0$, Q46-$\eta5$-$\zeta0$, and Q47-$\eta5$-$\zeta0$, respectively. This indicates that shocks are more frequent for smaller $Q_j$, consistent with the fact that more extended cocoons of backflow develop for lower power jets. The convergence of the PDFs in the default (thin lines) and high-resolution (thick lines) simulations for the Q46 and Q47 models, shown in panels (b) and (c), looks good. Yet, there are more shocks in higher resolution jets, $N_t/(N_{\rm shock}N_j)=0.32$ and 0.38 for the Q46-$\eta5$-$\zeta0$-H and Q47-$\eta5$-$\zeta0$-H models, respectively. This confirms the previous statement that finer structures and more flow motions develop at smaller scales in higher resolution simulations.

Although the time evolution of the PDF of $M_s$ is not shown here, we find that shocks are somewhat stronger at earlier times, in particular, at $t\lesssim10~t_{\rm cross}$. On the other hand, after $t\sim20~t_{\rm cross}$, the overall distribution of $M_s$ does not change significantly over the time period of our simulations, whereas the jet propagation speed fluctuates somewhat (see Figure \ref{f5}).

Panels (d)-(f) of Figure \ref{f9} show the PDFs of the shock speed, $N(\beta_s)N_j/N_t$, for shock zones in the different regions of the jet, where $\beta_{s} = v_{s}/c$. The speed is the largest for shocks in the jet spine flow and the next largest for shocks in the backflow; shocks in the jet spine flow are relativistic with $\beta_{s}\sim 0.2-1.0$, while those in the backflow are mildly or sub-relativistic with $\beta_{s}\sim 0.01-0.4$. On the other hand, the bow shock and shocks in the shocked ICM are non-relativistic with characteristic values of $\beta_{s}\lesssim 0.005-0.05$, which increase with increasing $Q_j$.

\begin{figure*}[t]
\centering
\vskip 0.1 cm
\includegraphics[width=1\linewidth]{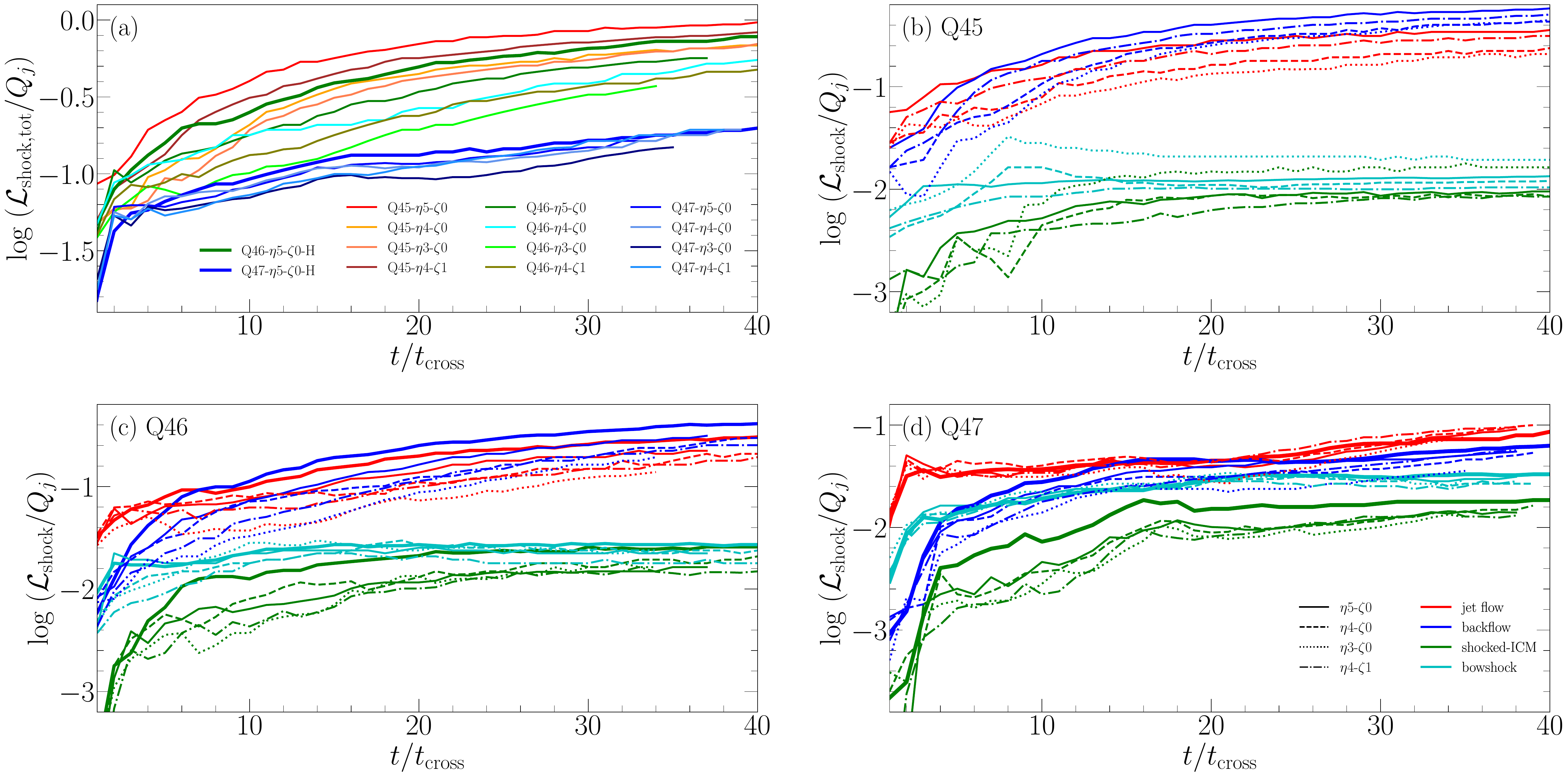}
\vskip -0.1 cm
\caption{Integrated energy dissipation rate at shocks with $M_s\geq1.01$, $\mathcal{L}_{\rm shock}$, normalized by $Q_j$, as a function of time. (a) Total dissipation rate, $\mathcal{L}_{\rm shock,tot}$, due to all the shocks for all the models listed in Table \ref{tab:t1}. The models with the same $Q_j$ are shown in the same hue of colors, reddish for Q45, greenish for Q46, and bluish for Q47. (b)-(d) $\mathcal{L}_{\rm shock}$ due to shocks in the jet spine flow (red), backflow (blue), shocked ICM (green), and bow shock (cyan) for the Q45, Q46, and Q47 models, respectively. The solid lines are used for the fiducial models with $\eta5$-$\zeta0$, the dashed lines for $\eta4$-$\zeta0$, the dotted lines for $\eta3$-$\zeta0$, and the dot-dashed lines for $\eta4$-$\zeta1$. The thick solid lines in panels (a), (c), and (d) are for high-resolution models.}
\label{f10}
\end{figure*}

\subsection{Shock Dissipation}
\label{s4.3}

To measure the importance of shocks in the jet flow dynamics, we quantify the shock dissipation by estimating the rate of the jet kinetic energy converted into heat at shocks. Panels (g)-(i) of Figure \ref{f9} show the energy dissipation rate at shock zones with the Mach number in the range of [$\log M_s$,$\log M_s+ d\log M_s$],
\begin{equation}
L_{\rm shock}(M_s) = \sum_{\log M_s}^{\log M_s+d\log M_s}\Delta\Phi_{\rm shock}A_{\rm shock},
\label{lke}
\end{equation}
which is normalized to the energy injection rate of the jet, $Q_j$. Here, $A_{\rm shock}=1.19(\Delta x)^2$ is the mean projected area of a shock zone within 3D space for random shock normal orientations \citep[e.g.,][]{ryu2003}. The plots demonstrate that shocks in the jet spine flow and backflow are dominant in the shock dissipation, while the bow shock, even with high $M_s$'s, is relatively unimportant. Despite much lower preshock densities, shocks in the jet spine flow and backflow has higher $\Delta\Phi_{\rm shock}$ than those in the shocked ICM and the bow shock, because they have much higher $v_s$.

In Figure \ref{f10}, we examine the integrated energy dissipation rate at shocks,
\begin{equation}
\mathcal{L}_{\rm shock}=\int_{M_{\rm min}} L_{\rm shock}(M_s)d\log M_s,
\end{equation}
as a function of time. Here, the minimum value of $M_s$ is $M_{\rm min}=1.01$. Note that $\mathcal{L}_{\rm shock}/Q_j$ is the fraction of the jet-injected energy, which is dissipated into heat at shocks. Panel (a) shows the total dissipation rate, $\mathcal{L}_{\rm shock,tot}/Q_j$ due to all the shocks in the jet-induced flows for all the models considered here, while panels (b)-(d) show $\mathcal{L}_{\rm shock}/Q_j$ due to shocks in the different regions for the Q45, Q46, and Q47 models, respectively. As the jet penetrates into the background medium, a cocoon filled with shocks and turbulence develops. This leads to the initial increase of the shock dissipation. After the jet-induced structures have more-or-less fully developed, for instance, at $t\gtrsim 20~t_{\rm cross}$, $\mathcal{L}_{\rm shock}/Q_j$ gradually approaches asymptotic values, although it seems to increase somewhat even close to $t_{\rm end}$ in our simulations.

The following points are noticeable in Figure \ref{f10}. (1) In panel (a), $\mathcal{L}_{\rm shock,tot}/Q_j$ is larger for smaller $Q_j$, as shocks are more frequent in lower power jets. However, still $\mathcal{L}_{\rm shock,tot}$ itself is larger for larger $Q_j$, since shocks on average have higher $v_s$ and dissipate larger amounts of energy in higher power jets. (2) Given the same $Q_j$, $\mathcal{L}_{\rm shock,tot}$ is somewhat smaller for the models with higher $\rho_j$ (smaller $\eta$). The increase of the jet pressure by an order of magnitude ($\zeta=10$) has only marginal effects on $\mathcal{L}_{\rm shock,tot}$. (3) In panels (b)-(d), the shock dissipation occurs mostly at shocks in the backflow (blue) and the jet spine flow (red), while the dissipation at shocks in the shocked ICM (green) and the bow shock (cyan) is much less. Thus, regarding the shock dissipation, the backflow is most important in the Q45 and Q46 models, while the backflow and the jet spine flow are about equally important in the Q47 models. (4) The comparison of $\mathcal{L}_{\rm shock}/Q_j$ for high-resolution models (bold-solid lines) with that for the corresponding default-resolution models (solid lines) demonstrates that the shock dissipation fraction is higher owing to more frequent shocks at higher resolution simulations.

As noted in Section \ref{s3.2}, the kinetic energy is not strictly defined in RHDs. We assume the form in Equation (\ref{Edecomp}), which reduces to the Newtonian kinetic energy in the non-relativistic limit. In addition, the kinetic energy dissipation rate at shocks is a frame-dependent quantity, and we employ $\Delta\Phi_{\rm shock}$ in Equation (\ref{keflux}), which is valid in the shock-rest frame. Thus, our estimations of $\mathcal{L}_{\rm shock,tot}$ should be considered to be only approximate. Nevertheless, $\mathcal{L}_{\rm shock,tot}/Q_j \sim 0.5 - 1$, $0.45-0.8$, and $0.1-0.15$ at $t=t_{\rm end}$ for the Q45, Q46, and Q47 models, respectively. These results indicate that a substantial fraction of the jet energy is dissipated at shocks in the jet-induced flows; shocks in the backflow is important for the shock dissipation, even though their speeds are mildly relativistic.

In diffusive shock acceleration (DSA) theory, cosmic rays (CRs) are scattered off MHD waves in the shock converging flows and gain energy by crossing repeatedly back and forth across the shock front \citep{bell1978}. In non-relativistic shocks, it predicts a power-law distribution of CRs, $f(p)\propto p^{-q}$, with slope, $q=4M_s^2/( M_s^2-1)$, in the test-particle regime \citep{drury1983}. The slope becomes $q=4$ for strong non-relativistic shocks with large $M_s$. In relativistic shocks, on the other hand, a power-low spectrum with a steeper slope, $q\sim 4.2-4.3$, is predicted \citep[e.g.,][]{achterberg2001}, and the acceleration becomes less efficient at the ultra-relativistic limit 
due to the anisotropic particle distribution and the limited residence time at both the upstream and downstream regions \citep[e.g.,][]{bykov2012,sironi2015}. Hence, among shocks in the jet-induced flows, we expect that mildly relativistic shocks with $v_s/c\lesssim 0.1$ in the backflow could play important roles in producing UHECRs, as previously suggested \citep{bell2018,matthews2019}.

On the other hand, as shown above, shocks are quite frequent with the mean distance between shock surfaces, $\langle l_{\rm sh}\rangle \sim0.3-0.6~r_j$, for shocks with $M_s \geq 1.01$ in the jet-induced flows. Considering $r_j=1$ kpc in our models, UHECRs with $E_{\rm CR} \gtrsim 1$ EeV may have, $r_g\approx 1.1~{\rm kpc} (E_{\rm CR}/10^{18}~{\rm eV}) (B/1 \mu {\rm G})^{-1}> \langle l_{\rm sh}\rangle$, that is, the gyroradius larger than the mean distance between shock surfaces. Hence, UHECRs may encounter more than one shocks in a scattering length and get accelerated differently, compared to a single episode of DSA. This rather complex scenario involving multiple shocks needs to be further examined.

\begin{figure*}[t] 
\centering
\vskip 0.1 cm
\includegraphics[width=0.9\linewidth]{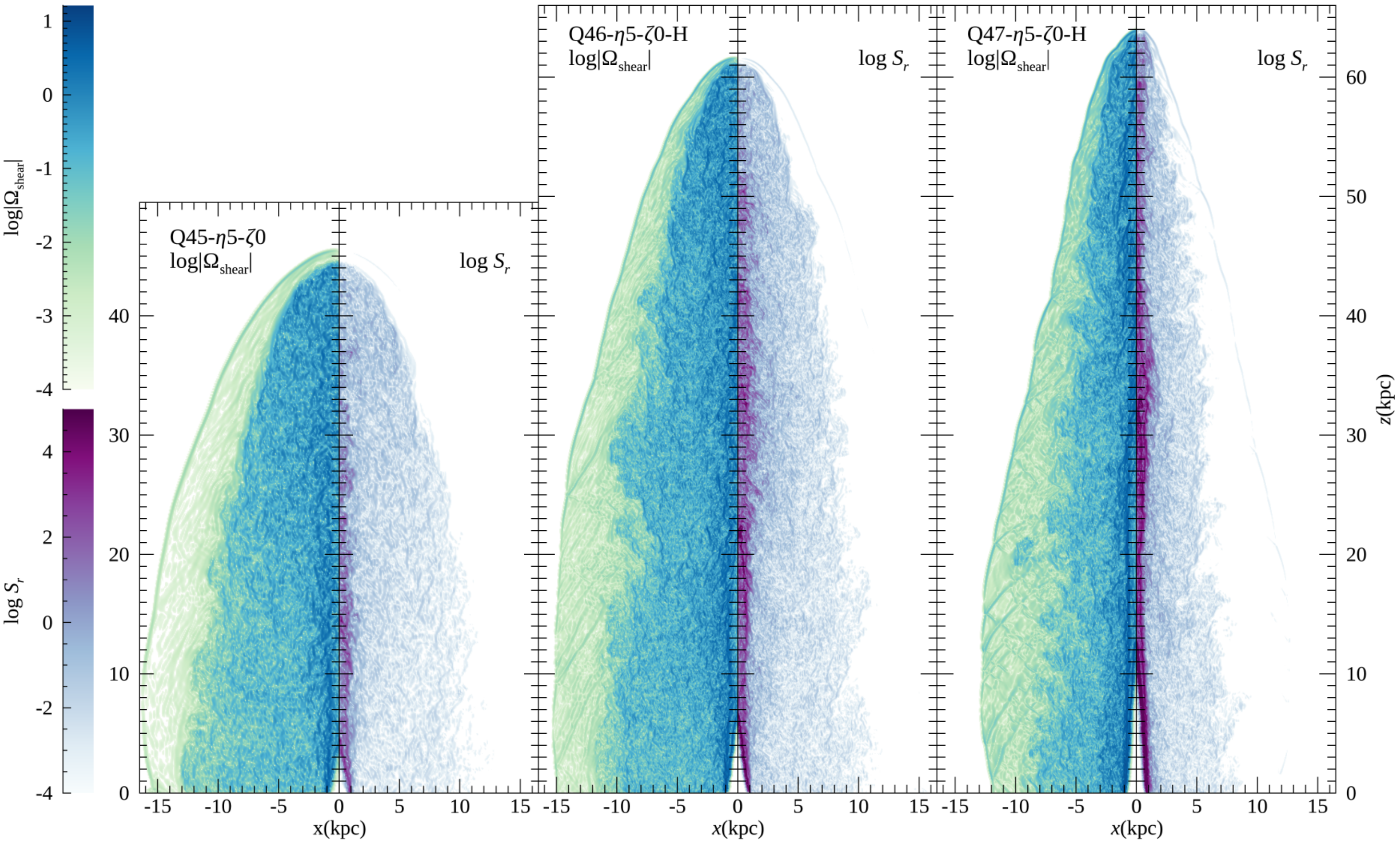}
\vskip -0.1 cm
\caption{2D slice images of the magnitude of the velocity shear, $\Omega_{\rm shear}\equiv|{\partial v_{z}}/{\partial r}|$, on the left side ($x<0$) and the relativistic shear coefficient, $\mathcal{S}_{r}$, on the right side ($x>0$) through the plane of $y=0$ for the three fiducial models, Q45-$\eta5$-$\zeta0$, Q46-$\eta5$-$\zeta0$-H, and Q47-$\eta5$-$\zeta0$-H, at $t=t_{\rm end}$. The velocity shear and shear coefficient are given in units of $c/r_j$ and $(c/r_j)^{2}$, respectively. See Figure \ref{f2} for the description of the axial ratio of the simulation domain.}
\label{f11}
\end{figure*}

Although not shown as plots, the comparison of the higher $\rho_j$ models ($\eta3$ and $\eta4$) with the fiducial models ($\eta5$) indicates that for the same $Q_j$, $N(M_s)$ for the first recollimation shock including the peak shifts to higher $M_s$ with higher $\rho_j$, since $p_j/\rho_j$ is lower and hence $c_{s,1}$ is smaller in the injected jet material. In addition, $N(\beta_s)$ for shocks in the jet spine flow and backflow shifts to lower $\beta_{s}$ with higher $\rho_j$ (smaller $v_j$ and $\Gamma_j$). Comparing the $\eta4$-$\zeta1$ models with the $\eta4$-$\zeta0$ models, in the over-pressured jets, $N(M_s)$ for the first recollimation shock shifts to lower $M_s$, owing to higher $c_{s,1}$ in the injected material. Otherwise, other characteristics of the PDFs seem to be less affected by different $\eta$ and $\zeta$.

\begin{figure*}[t] 
\centering
\vskip 0.1 cm
\includegraphics[width=1.0\linewidth]{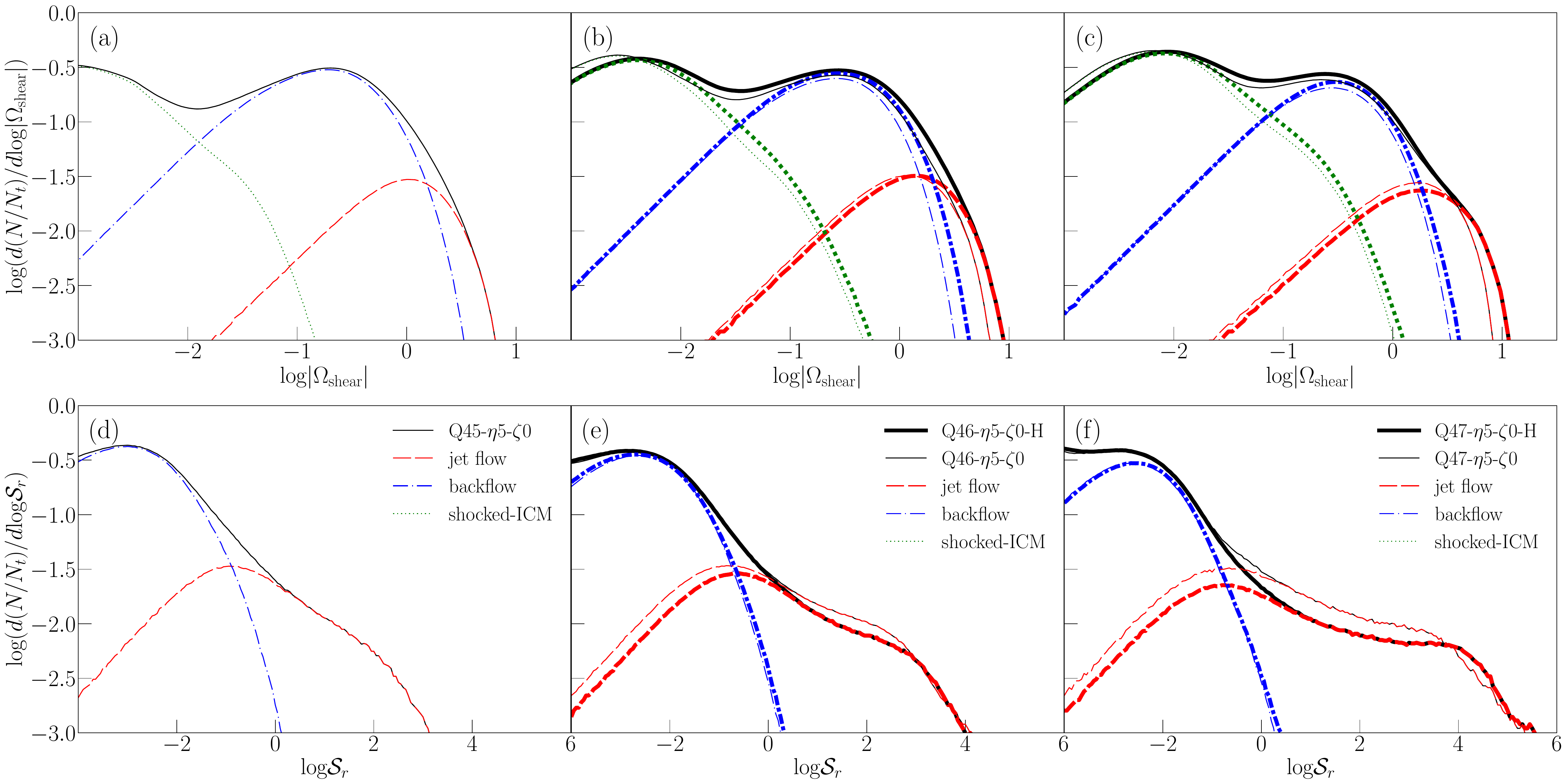}
\vskip -0.1 cm
\caption{PDFs of the magnitude of the velocity shear, $\Omega_{\rm shear}\equiv|{\partial v_{z}}/{\partial r}|$, (top panels) and the shear coefficient, $\mathcal{S}_r$, (bottom panels), averaged over $t/t_{\rm cross}=[20,30]$, for the jet spine flow (red dashed lines), backflow (blue dot-dashed lines), and shocked ICM (green dotted lines), for the five fiducial models listed in Table \ref{tab:t1}. The shear coefficient in the shocked ICM is very small and not shown. The black solid lines plot the PDFs for all the regions. The thick lines in the center and right panels are for high-resolution models. The velocity shear and shear coefficient are given in units of $c/r_j$ and $(c/r_j)^{2}$, respectively, and $N_t$ is the total number of grid zones in the volume encompassed by the bow shock surface.}
\label{f12}
\end{figure*}

\subsection{Properties of Velocity Shear}
\label{s4.4}

As shown in the left half of each panel in Figure \ref{f3}, a relativistic velocity shear with $\Delta v_z \sim c$ develops at the interface between the upward-moving jet spine flow (red) and the downward-moving backflow (blue). In addition, a non-relativistic velocity shear with $\Delta v_z < 0.1c$ develops at the interface between the backflow and the shocked ICM (white). These shear interfaces are unstable against the KH instability, so turbulence emerges in the jet spine flow and backflow as well as in the shocked ICM.

Assuming the presence of turbulent magnetic fields, UHECRs may be accelerated through elastic scatterings off turbulent waves moving with the underlying shear flow, especially around the jet-backflow boundary \citep[see, e.g.,][for a review]{rieger2019}. In the so-called {\it discrete shear acceleration}, which operates when the particle scattering length is larger than the width of the velocity shear layer, the mean energy gain is given as $\Delta E/E \sim \Gamma_j-1$ for each crossing of the jet-backflow boundary, if the CR distribution is nearly isotropic around the boundary \citep[e.g.,][]{ostrowski1998}. On the other hand, when the particle scattering length is smaller than the velocity shear scale, the so-called {\it gradual shear acceleration} operates and the energy gain is $\Delta E/E \propto (\bar{v}/c)^2$ (Fermi-II), where $\bar{v}=(\partial v_z/\partial r)\lambda$ is the effective velocity difference that the particles with the mean free path, $\lambda$, experience in the shear flow with $\mbox{\boldmath$v$}= v_z(r)\hat{\mbox{\boldmath$z$}}$ \citep{rieger2019}. With the relativistic gradual shear of the jet, the acceleration timescale is inversely proportional to the relativistic shear coefficient defined as,
\begin{equation}
\mathcal{S}_{r} \equiv \frac{\Gamma_{z}^{4}}{15}\left(\frac{\partial v_{z}}{\partial r}\right)^{2},
\label{relshearcoeff}
\end{equation}
where $\Gamma_{z}=1/\sqrt{1-(v_{z}/c)^{2}}$ \citep{webb2018}.

Figure \ref{f11} shows the 2D slice images of the magnitude of the velocity shear, $\Omega_{\rm shear} \equiv |{\partial v_{z}}/{\partial r}|$ (in the left half of each panel, $x<0$), and the relativistic shear coefficient, $\mathcal{S}_{r}$ (in the right half of each panel, $x>0$), for three fiducial models. Figure \ref{f12} plots their PDFs, $N(\Omega_{\rm shear})$ (top panels) and $N(\mathcal{S}_{r})$ (bottom panels), in the different regions for five fiducial models. The value of $\mathcal{S}_{r}$ in the shocked ICM is very small, so its PDF is not shown. While the velocity shear is somewhat larger for higher power jets, the dependence on $Q_j$ is not strong. The peaks of $N(\Omega_{\rm shear})$ lie at $\Omega_{\rm shear} (r_j/c)\gtrsim 1$ for the jet spine flow, $\sim 0.1-1$ for the backflow, and $\lesssim 0.01$ for the shocked ICM, respectively. On the other hand, $\mathcal{S}_{r}$ in the jet spine flow is noticeably larger for higher power jets, as can be seen with the purple tone shade in Figure \ref{f11}, owing to the weighting factor of $ \Gamma_{z}^4$. Figure \ref{f12} shows that it extends up to $\mathcal{S}_{r} (r_j/c)^2\sim 10^3-10^5 $ inside the jet spine flow (red dashed lines), depending on $Q_j$. In the backflow, $\mathcal{S}_{r}$ depends only weakly on $Q_j$, and $N(\mathcal{S}_{r})$ peaks at $\mathcal{S}_{r}(r_j/c)^2 \sim 10^{-3}-10^{-2}$. Although not shown here, $\Omega_{\rm shear}$ and $\mathcal{S}_{r}$ are larger in the earlier stage, but their PDFs converge as the jet-induced structures approach more or less steady states at $t\gtrsim 20~ t_{\rm cross}$. Again although not shown here, the models with the same $Q_j$ but different $\eta$ and $\zeta$ have similar $N(\Omega_{\rm shear})$ and $N(\mathcal{S}_{r})$, except that $\mathcal{S}_{r}$ is smaller in the models with higher $\rho_j$ (smaller $\Gamma_j$).

While both $|{\partial v_{z}}/{\partial r}|$ and $\mathcal{S}_{r}$ are the largest in the jet spine flow, the cocoon of the backflow occupies a much larger volume than the jet spine. Hence, we expect that lower energy CRs would be accelerated via the gradual shear acceleration mostly in the backflow \citep[e.g.,][]{rieger2004}. In comparison, higher energy CRs with the gyroradius $\gtrsim 1$ kpc may diffuse across the jet-backflow boundary and gain energy via the discrete shear acceleration \citep[e.g.,][]{kimura2018}. However, the exact process of the shear acceleration will depend on the details such as the jet geometry, the origin of seed particles, the magnetic field strength and configuration, the magnetic fluctuations that scatter CRs, and etc.

\subsection{Properties of Vorticity}
\label{s4.5}

\begin{figure*}[t] 
\centering
\vskip 0.1 cm
\includegraphics[width=0.9\linewidth]{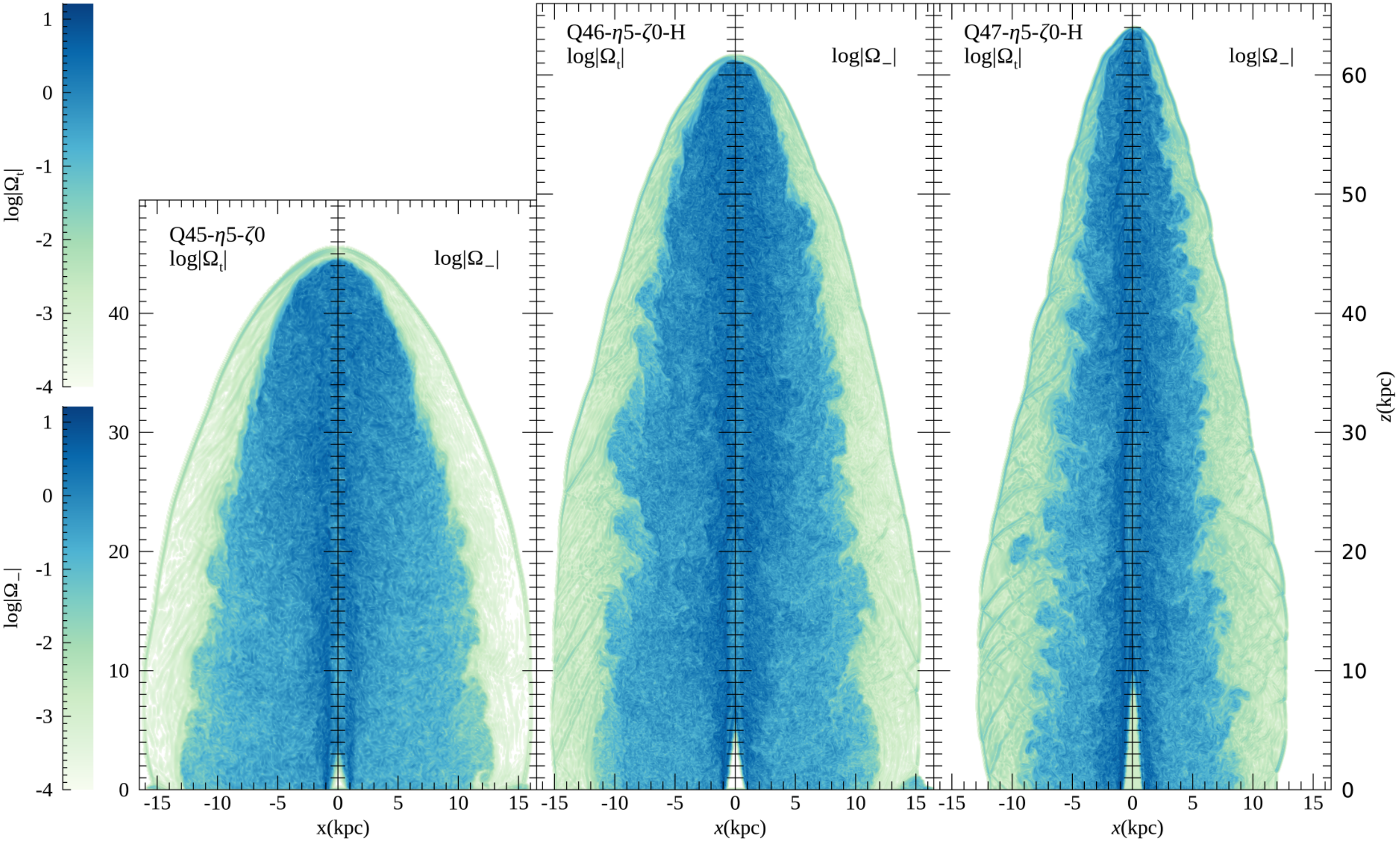}
\vskip -0.1 cm
\caption{2D slice images of the magnitudes of the total vorticity, $\Omega_t$, on the left side ($x<0$) and the vorticity excluding the shear, $\Omega_-$, on the right side ($x>0$) through the plane of $y=0$ for the three fiducial models, Q45-$\eta5$-$\zeta0$, Q46-$\eta5$-$\zeta0$-H, and Q47-$\eta5$-$\zeta0$-H, at $t=t_{\rm end}$. See the text for the definitions of $\Omega_t$ and $\Omega_-$. The vorticity is given in units of $c/r_j$. See Figure \ref{f2} for the description of the axial ratio of the simulation domain.}
\label{f13}
\end{figure*}

\begin{figure*}[t] 
\centering
\vskip 0.1 cm
\includegraphics[width=1\linewidth]{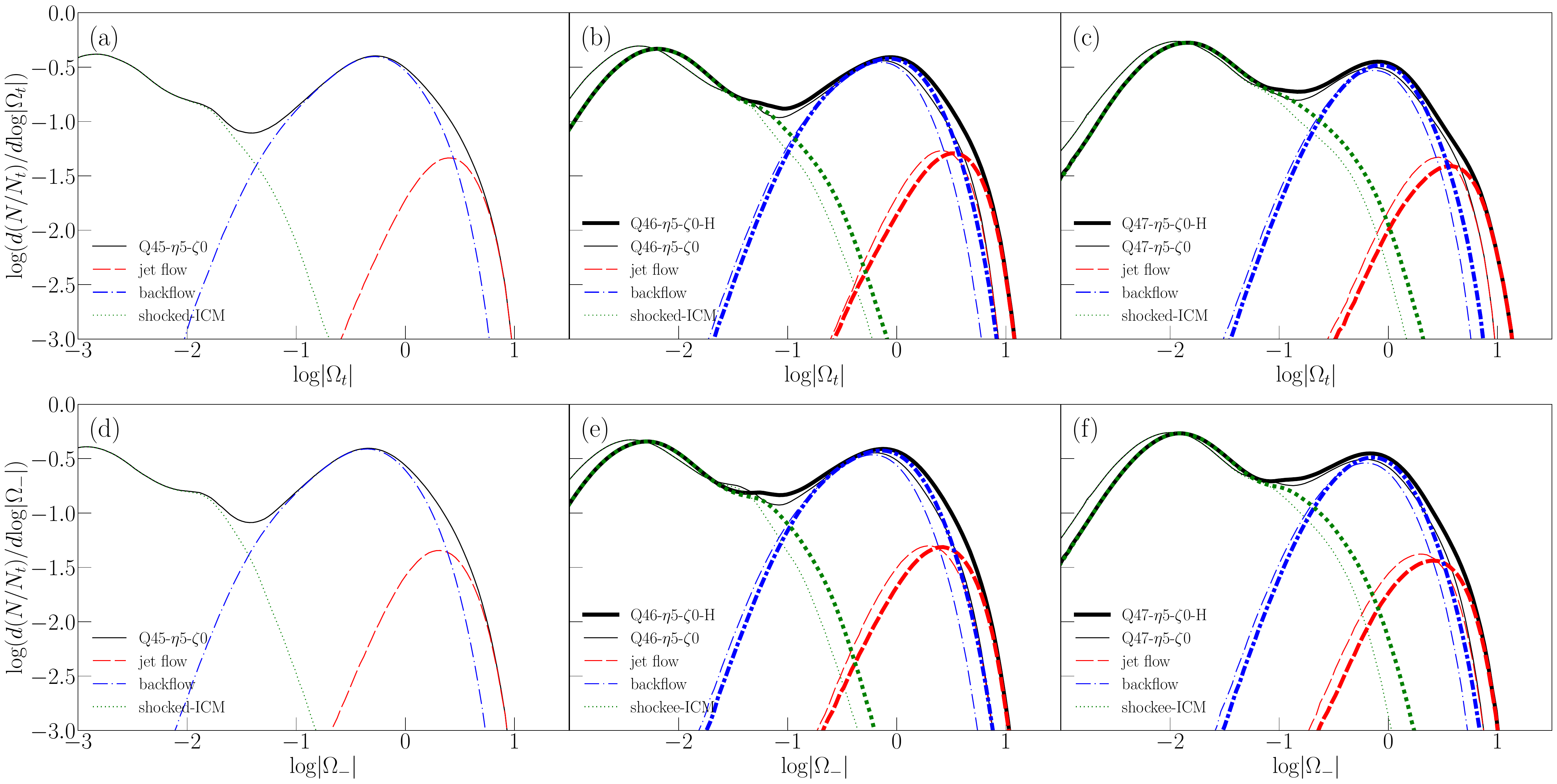}
\vskip -0.1 cm
\caption{PDFs of the magnitudes of the total vorticity, $\Omega_t$, (top panels) and the vorticity excluding the shear, $\Omega_-$, (bottom panels), averaged over $t/t_{\rm cross}=[20,30]$, for the jet spine flow (red dashed lines), backflow (blue dot-dashed lines), and shocked ICM (green dotted lines), for the five fiducial models listed in Table \ref{tab:t1}. See the text for the definitions of $\Omega_t$ and $\Omega_-$. The black solid lines plot the PDFs for all the regions. The thick lines in the center and right panels are for high-resolution models. The vorticity is given in units of $c/r_j$, and $N_t$ is the total number of grid zones in the volume encompassed by the bow shock surface.}
\label{f14}
\end{figure*}

As discussed above, turbulence develops in the jet-induced flows as a consequence of the KH instability in the shear boundaries and also the ensuing complex flow dynamics. 
Then, along with the shear acceleration, the stochastic Fermi-II acceleration due to elastic scatterings off turbulent magnetic fluctuations should operate, especially inside the cocoon \citep[e.g.,][]{osullivan2009,hardcastle2010}. 
As a measure of the turbulence, we define the vorticity excluding the shear considered in the previous subsection as
\begin{equation}
\mbox{\boldmath$\Omega$}_- = \mbox{\boldmath$\Omega$}_t +{{\partial v_{z}}\over {\partial r}}\hat{\mbox{\boldmath$\theta$}}, 
\end{equation}
where $\mbox{\boldmath$\Omega$}_t=\mbox{\boldmath$\nabla$}\times\mbox{\boldmath$v$}$ is the total vorticity.

Figure \ref{f13} shows the 2D slice images of $\Omega_t = |\mbox{\boldmath$\Omega$}_t|$ (in the left half of each panel, $x<0$) and $\Omega_- = |\mbox{\boldmath$\Omega$}_-|$ (in the right half of each panel, $x>0$) for three fiducial models. The spatial distributions of $\Omega_t$ and $\Omega_-$ are similar to those of the velocity shear, $\Omega_{\rm shear}$, shown in Figure \ref{f11}. They look more or less homogeneous in the cocoon, as expected in fully developed turbulence, but patterns appear in the shocked ICM. Some of the patterns follow the shocks shown in Figure \ref{f8}, and others match the herringbone patterns in the density distributions in Figure \ref{f3}. While those patterns look interesting, overall the flow dynamics in the shocked ICM is not very important in terms of the energy dissipations, as shown in Figure \ref{f10} and also in Figure \ref{f16} below.

Figure \ref{f14} plots the PDFs of $\Omega_t$ and $\Omega_-$, $N(\Omega_t)$ (top panels) and $N(\Omega_-)$ (bottom panels), for five fiducial models. They behave similarly as $N(\Omega_{\rm shear})$ shown in Figure \ref{f12}; the mean values of $\Omega$'s are the greatest in the jet spine flow (red lines) and the smallest in the shocked ICM (green lines). Moreover, they show only weak dependence on the jet power, $Q_j$, although the vorticities increase slightly with increasing $Q_j$. We note that on average $\Omega_-$ is comparable to, or somewhat larger than, $\Omega_{\rm shear}$, and $N(\Omega_-)$ peaks at slightly higher values than $N(\Omega_{\rm shear})$. This indicates that turbulence develops fully in the shear boundaries, and hence the radial and vertical components of $\mbox{\boldmath$\Omega$}_t$ is comparable to $\Omega_{\rm shear}$. As in the case of $\Omega_{\rm shear}$, both $\Omega_t$ and $\Omega_-$ are larger in the earlier stage, but their PDFs converge at $t\gtrsim 20~ t_{\rm cross}$. Furthermore, the models with the same $Q_j$ but different $\eta$ and $\zeta$ have similar PDFs.

Our finding, $\Omega_- \gtrsim \Omega_{\rm shear}$, implies that the stochastic turbulent acceleration of UHECRs could be as important as the gradual shear acceleration for CRs with the diffusion length smaller than the width of the velocity shear layers of the jet. However, the real situation could be quite complex, since the two acceleration processes may operate simultaneously and their interplay would occur. Then, they may need to be examined as the turbulent shear acceleration \citep[e.g.,][]{ohira2013}.

\begin{figure*}[t] 
\centering
\vskip 0.1 cm
\includegraphics[width=1\linewidth]{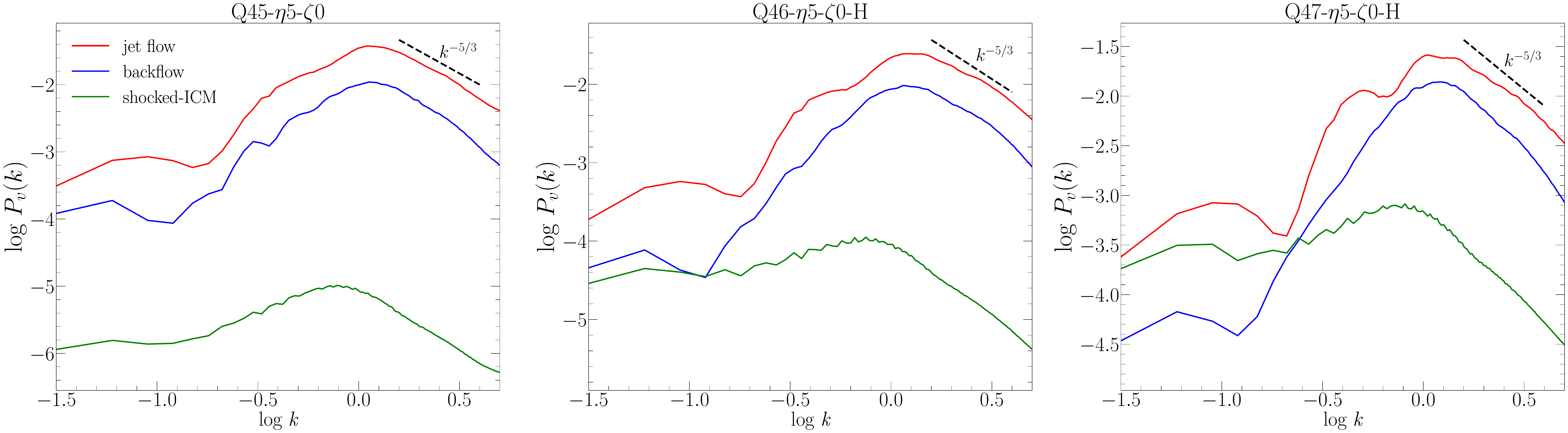}
\vskip -0.1 cm
\caption{Power spectra of the fluid three-velocity of turbulent flows, $\mbox{\boldmath$v$}_{\rm turb}$, averaged over $t/t_{\rm cross}=[20,30]$, in the jet spine flow (red lines), backflow (blue lines), and shocked ICM (green lines), for the three fiducial models, Q45-$\eta5$-$\zeta0$, Q46-$\eta5$-$\zeta0$-H, and Q47-$\eta5$-$\zeta0$-H. They are normalized in the way that $\int P_v(k) dk = v_{\rm rms}^2/c^2$ of turbulent flows in each region. The wavenumber $k=1$ corresponds to the wavelength $\lambda=r_j$.}
\label{f15}
\end{figure*}

\subsection{Turbulent Dissipation}
\label{s4.6}

The turbulence developed in the jet-induced flows cascades down to smaller scales and is eventually converted into heat. Here, we examine this process, known as the turbulent dissipation, to measure the importance of turbulence in the jet flow dynamics and also to assess the relative importance of the turbulent versus shock dissipations in relativistic jets.

Despite the relevance to many astrophysical phenomena such as relativistic jets and accretion disks, turbulence in relativistic regime has been less explored than its non-relativistic counterpart. Recently, however, systematic studies of RHD and RMHD turbulence using high-resolution simulations have been performed \citep[e.g.,][]{zrake2012,zrake2013,radice2013}. In particular, the RHD studies showed that the velocity power spectrum follows the $k^{-5/3}$ power-law and the kinetic energy dissipation scaled as $v_{\ell}^3/\ell$ is independent of $\ell$, in good agreement with the predictions of the classical Kolmogorov theory, at least, for mildly relativistic, hydrodynamic flows with the mean Lorentz factor of $\lesssim$ a few \citep[see, e.g.,][]{zrake2013,radice2013}. Here, $k$ is the wavenumber and $v_{\ell}$ is the characteristic speed of turbulent flows at the length scale, $\ell$.

Our simulated jets contain the bulk motions of the upward-moving jet spine flow and the downward-moving backflow. Hence, the flow motions associated with turbulence need to be separated from the large-scale bulk motions. The extraction of turbulent flow velocity could be done by applying a ``filtering'', and we employ the scheme used in \citet{vazza2017}. The mean velocity is computed as
\begin{equation}
\left<\mbox{\boldmath$v$}(\mbox{\boldmath$r$})\right>_\Lambda=\frac{\sum_i w_i \mbox{\boldmath$v$}_i}{\sum_i w_i},
\label{turbfilt}
\end{equation}
where the summation runs over the cubic box of the size $\Lambda$ around the position $\mbox{\boldmath$r$}$. Here, $w_i$ is a weighting function, and we set $w_i=1$. Then, the turbulent velocity is estimated as $\mbox{\boldmath$v$}_{\rm turb}(\mbox{\boldmath$r$}) = \mbox{\boldmath$v$}(\mbox{\boldmath$r$})-\left<\mbox{\boldmath$v$}(\mbox{\boldmath$r$})\right>_\Lambda$. The estimated $\mbox{\boldmath$v$}_{\rm turb}$ should depend on the filtering size, $\Lambda$. \citet{vazza2017} suggested that the mean size of eddies in turbulent flow motions would be a good choice for $\Lambda$.  Here, we set $\Lambda=r_j$, since the jet radius represents a characteristic scale in the jet-induced flows.

Figure \ref{f15} shows the power spectrum, $P_v$, of $\mbox{\boldmath$v$}_{\rm turb}$ in the different regions for three fiducial models. A few points are evident. (1) The peaks of $P_v$ occur around $k\sim 1$, corresponding to the wavelength $\lambda \sim r_j$, due to the filtering scale we impose. (2) At $k\gtrsim 1$, i.e., at smaller scales of $\lambda< r_j$, $P_v$ exhibits the Kolmogorov power-law of $k^{-5/3}$, despite the fact that there are numerous shocks in the flows. If the velocity spectrum is dominated by shocks in non-relativistic regime, $P_v$ follows $k^{-2}$, the Burger's power spectrum \citep[e.g.,][]{kim2005}. The Kolmogorov spectrum here should reflect the characteristics of RHD turbulence \citep{zrake2013,radice2013}. In addition, the contribution of shocks to $P_v$ may not be substantial, because most shocks are weak with $M_s \lesssim$ a few. (3) $P_v$ is the largest in the jet spine flow, the next largest in the backflow, and the smallest in the shocked ICM, revealing the rms magnitude of turbulent flow motions. In addition, although not shown here, the solenoidal mode, $P_{\rm sol}$, dominates over the compressive mode, $P_{\rm comp}$, in the jet spine flow and backflow; $P_{\rm sol}$ is about an order of magnitude larger than $P_{\rm comp}$ there. On the other hand, in the shocked ICM, while $P_{\rm sol}$ is still larger than $P_{\rm comp}$, the difference of the two is small.

\begin{figure*}[t] 
\centering
\vskip 0.1 cm
\includegraphics[width=1\linewidth]{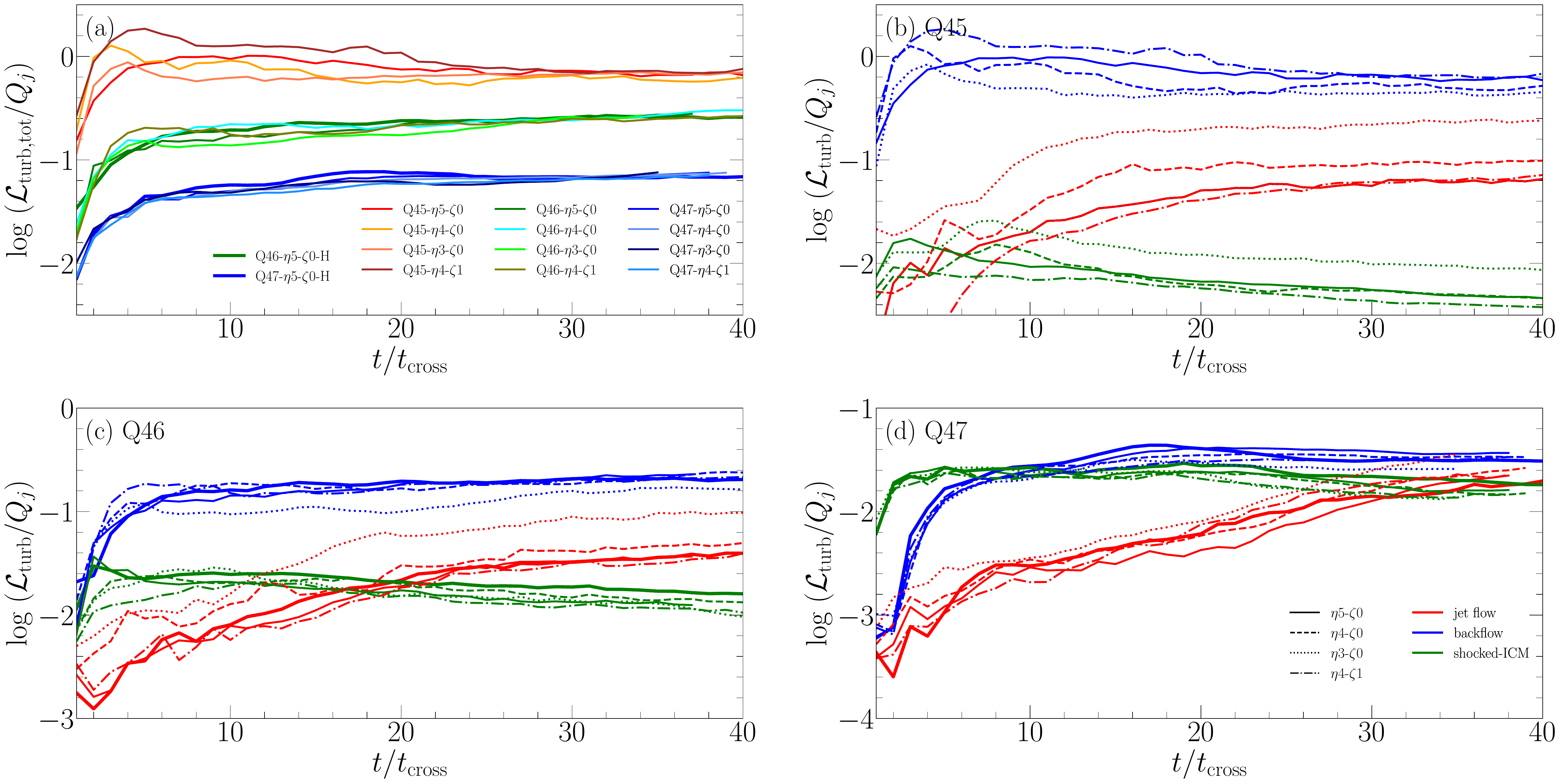}
\vskip -0.1 cm
\caption{Energy dissipation rate through turbulent cascade as a function of time. (a) Total dissipation rate, $\mathcal{L}_{\rm turb,tot}$, for all the models listed in Table \ref{tab:t1}. The models with the same $Q_j$ are shown in the same hue of colors, reddish for Q45, greenish for Q46, and bluish for Q47. (b)-(d) $\mathcal{L}_{\rm turb}$ due to the turbulence in the jet spine flow (red), backflow (blue), and shocked ICM (green) for Q45 models (b), Q46 models (c), and Q47 models (d), respectively. The thick lines in (a), (c), and (d) are for high-resolution models. The turbulent dissipation is normalized by the jet energy, $Q_j$.}
\label{f16}
\end{figure*}

We then estimate the energy dissipation rate due to turbulent flow motions, $\mathcal{L}_{\rm turb}$, using the following approximate quantity:
\begin{equation}
\mathcal{L}_{\rm turb} \approx  \varphi \frac{E_{\rm turb}}{\tau_{\rm cascade}},
\label{turbdissp}
\end{equation}
where $\varphi$ is a free parameter of order of unity and to be set as $\varphi=1$. Here, $E_{\rm turb}\approx \Gamma_{\rm turb}\left(\Gamma_{\rm turb}-1\right)\rho c^2$ is the kinetic energy of turbulent flows, where $\Gamma_{\rm turb} = 1/\sqrt{1-(v_{\rm turb}/c)^2}$. And $\tau_{\rm cascade} \approx \lambda_{\rm peak}/v_{\rm rms}$ is the timescale of turbulent cascade, where $v_{\rm rms}$ is the rms of $\mbox{\boldmath$v$}_{\rm turb}$ and $\lambda_{\rm peak}\sim r_j$ is the peak scale of $P_v$. Then, $\mathcal{L}_{\rm turb}/Q_j$ can be regarded as the fraction of the jet-injected energy, dissipated into heat through turbulent cascade.
  
In Figure \ref{f16}, panel (a) shows the total $\mathcal{L}_{\rm turb,tot}/Q_j$ due to turbulent flow motions inside the entire volume of the jet-induced structure as a function of time for all the models, while panels (b)-(d) show $\mathcal{L}_{\rm turb}/Q_j$ in the different regions for the Q45, Q46, and Q47 models, respectively. After the initial quick increase during $t\lesssim$ several $\times~t_{\rm cross}$, $\mathcal{L}_{\rm turb,tot}/Q_j$ approaches asymptotic values.

The following points can be learned from Figure \ref{f16}. (1) As in the case of the shock dissipation, $\mathcal{L}_{\rm turb,tot}/Q_j$ is larger for jets with smaller $Q_j$, because more extensive cocoons filled with turbulence develop in lower power jets. However, still $\mathcal{L}_{\rm turb,tot}$ itself is larger for jets with larger $Q_j$, because turbulent flows are on average more energetic in higher power jets. (2) While the total $\mathcal{L}_{\rm turb,tot}/Q_j$ is almost independent of $\eta$ and $\zeta$, $\mathcal{L}_{\rm turb}/Q_j$ in the separate regions does depend on those parameters. For instance, in the models with higher density, the relative contribution from the jet spine flow is larger, while that from the backflow is smaller. (3) In panels (b)-(d), the turbulent dissipation is the largest in the backflow (blue) in the Q45 and Q46 models. For the Q47 models with smaller cocoons, the contributions from all the three regions seem to be important. In comparison, in the case of $\mathcal{L}_{\rm shock}$, shocks in the backflow and the jet spine flow make similar contributions. (4) The comparison of $\mathcal{L}_{\rm turb}/Q_j$ for high-resolution models (bold-solid lines) with that for the corresponding default-resolution models (solid lines) indicates that $\mathcal{L}_{\rm turb}$ is better resolution-converged than $\mathcal{L}_{\rm shock}$. This is because with the Kolmogorov power spectrum, $v_r^3/r$ is independent of the length scale, and hence the estimate of the turbulent dissipation is not much affected by small scale flow motions.

The kinetic energy used in Equation (\ref{turbdissp}) is not strictly valid in RHDs as the shock kinetic energy dissipation in Equation (\ref{keflux}). The filtering with $\left<\mbox{\boldmath$v$}(\mbox{\boldmath$r$})\right>_\Lambda$ in Equation (\ref{turbfilt}), adopted to extract the turbulent component of flow motions, represents only one form out of many possibilities. In addition, the fuzzy factor $\varphi$ in Equation (\ref{turbdissp}) for the turbulent energy dissipation rate is not known \citep[e.g.,][]{maclow1999}. Thus, our estimations of $\mathcal{L}_{\rm turb}$ should be considered only approximate. Still, with the asymptotic values of $\mathcal{L}_{\rm turb,tot}/Q_j \sim 0.07-0.65$ in Figure \ref{f16}(a), we argue that the turbulent dissipation would be important for the flow dynamics in our simulated jets, especially in low-power jets.

Combing the turbulent dissipation with the shock dissipation in Figure \ref{f10}, we get $(\mathcal{L}_{\rm shock,tot}+\mathcal{L}_{\rm turb,tot})/Q_j \gtrsim 1$, $\lesssim 1$, and $< 1$ for the Q45, Q46, and Q47 models, respectively. Note that the dissipations by shocks and turbulence are the two channels through which the energy injected by the jet is converted into heat in RHDs. Hence, $(\mathcal{L}_{\rm shock}+\mathcal{L}_{\rm turb})/Q_j \gtrsim 1$ for the Q45 models indicates that our estimates may not be exact. Nevertheless, it implies that the low-power jets of the Q45 models have reached a steady state in the sense that the injection of the jet energy is about balanced with the dissipation. On the other hand, the higher power jets for the Q46 and Q47 models are still dynamically evolving in our simulations.

Overall, $\mathcal{L}_{\rm shock,tot}$ is somewhat larger than $\mathcal{L}_{\rm turb,tot}$ in our estimations. However, considering the uncertainties mentioned above, we suggest that the two types of dissipation would be comparable, which was also argued for in simulation studies of turbulence in non-relativistic regime \citep[e.g.,][]{park2019}. Hence, we conclude that both shocks and turbulence are important for the jet dynamics, and could play significant roles in the production of UHECRs in radio galaxies. The relative importance of different acceleration processes such as DSA, shear acceleration, and stochastic turbulent acceleration, however, needs be investigated further through detailed numerical studies in realistic jet flows.

\section{Summary}
\label{s5}

\begin{deluxetable}{cccc}[t]
\tablecaption{Properties in Different Regions$^a$\label{tab:t2}}
\tablenum{2}
\tabletypesize{\small}
\tablecolumns{4}
\tablewidth{0pt}
\tablehead{
\colhead{Parameters} & 
\colhead{jet spine} & 
\colhead{backflow}& 
\colhead{shocked-ICM}
}
\startdata
$\beta_{s}=v_{s}/c$ & $\sim 0.4$ & $\sim 0.06-0.2$ & $\sim 0.01$\\
$\mathcal{L}_{\rm shock}/Q_j$ & $\sim 0.3$ & $\sim 0.4$ & $\sim0.03$\\
$\Omega_{\rm shear}/({c / r_j})$ & $\sim1.5$ & $\sim0.3$ & $\sim0.004$\\
$S_r/({c / r_j})^2$ & $\sim 0.2 $ &  $\sim 0.004$ & - \\
$\Omega_-/({c / r_j})$ & $\sim2.5$ & $\sim0.7$ & $\sim0.005$\\
$\mathcal{L}_{\rm turb}/Q_j$ & $\sim 0.04$ & $\sim 0.2$ & $\sim0.02$\\
\enddata
\tablenotetext{a}{Characteristic values are given for the high-resolution Q46 model, Q46-$\eta5$-$\zeta0$-H. Note that these  quantities have very broad distributions and depend on the jet model parameters.}
\end{deluxetable}

In Paper I, we have developed a novel RHD code, equipped with a high-order accurate shock-capturing scheme, WENO-Z, and a high-order accurate stability-preserving time discretization method, SSPRK, along with a realistic EOS that closely emulates the thermodynamics of relativistic perfect gas, RC. Using this code, we have performed a set of RHD simulations for relativistic, light jets with the initial bulk Lorentz factor $\Gamma_j\approx 2-70$, intending to study FR-II radio galaxies on $\sim 50$ kpc scales. The uniform background medium is defined by the parameters relevant for the hot ICM, that is, $n_{\rm H,ICM} \approx 10^{-3}$ cm$^{-3}$ and $T_{\rm ICM} \approx 5\times10^7$ K. The model jets are specified by several model parameters: the jet power, $Q_j\approx 3\times 10^{45} - 3\times 10^{47} {\rm erg ~s^{-1}}$, the jet-to-background density contrast, $\eta =10^{-5} - 10^{-3}$,  and the pressure contrast, $\zeta=1-10$ (see Table \ref{tab:t1}). Here, we do not consider lower power FR-I type jets with $Q_j\lesssim 10^{45} {\rm erg ~s^{-1}}$, where the entrainment and mass-loading on subgrid scales as well as the dissipation through small-scale instabilities are expected to be important. 

As illustrated in Figure \ref{f1}, the jet-induced structures can be differentiated as follows: (1) the upward-moving jet spine flow with low density, relativistic adiabatic index $\gamma\approx 4/3$, and relativistic bulk flow speed $v\sim c$, (2) the downward-moving backflow with low density, mildly relativistic $\gamma$, and mildly relativistic $v\sim 0.2c$, (3) the non-relativistic shocked ICM with high density and $\gamma\approx 5/3$, and (4) the bow shock plowing through the dense background ICM \citep[e.g.,][]{marti2019}. The interfaces of the jet-backflow and the backflow-ICM have velocity shears and are unstable against the KH instability. As a result, chaotic turbulent structures, including an inflated cocoon, develop, instead of a stable working surface (termination shock) and ordered shear discontinuities.

The primary parameter that controls the jet morphology is the jet power $Q_j$ (see Fig. \ref{f2}), as pointed in previous studies \citep[e.g.][]{li2018}. The jet with higher $Q_j$ (or higher $\Gamma_j$) has a faster advance speed $v_{\rm head}$, and generates a more elongated cocoon. On the other hand, the jet with lower $Q_j$ penetrates more slowly into the background medium, and produces a more extended cocoon filled with shocks and turbulence. The secondary parameters, $\eta$ and $\zeta$, play less significant roles (see Fig. \ref{f3}). 

Utilizing the high-resolution capability of our new RHD code, we have examined the properties of nonlinear flow dynamics such as shocks, velocity shear, and turbulence in the jet-induced structures. The physical quantities associated with them in the different regions of the jet have rather broad distributions. Table \ref{tab:t2} lists the values at the peaks of distributions for the Q46-$\eta5$-$\zeta0$-H model with $Q_j= 3.34\times 10^{46}$ as ``characteristic values'', to provide only a general overview of the relevant physical quantities. Note that while they depend on $Q_j$, the dependence on $\eta$ and $\zeta$ is relatively weak.

{\it Shocks:} In the jet-induced flows, two types of shocks form. The bow shock and recollimation shocks show visually distinct surfaces. On the other hand, shocks induced in turbulent flows such as the jet spine flow, backflow, and shocked ICM have disordered surfaces of smaller sizes. Those shock surfaces are composed of many shock zones with varying parameters such as $M_s$ and $v_s$. The PDF analysis of $M_s$ of shock zones indicates that the bow shock has the characteristic Mach number $M_{s} \sim 3-13$, which is larger for higher $Q_j$. In contrast, the characteristic Mach number of the first recollimation shock is almost independent of $Q_j$ and is $M_{s} \sim 3$ for our fiducial models, but it is larger if $p_j/\rho_j$ is smaller and hence the preshock sound speed, $c_{s,1}$, is smaller. On the other hand, the PDFs of $M_s$ for shocks in turbulent flows have power-law forms. Shocks in the jet spine flow have the relativistic speeds of $\beta_s \sim 0.2 - 1$ and $M_{s}\lesssim 5$, while those in the backflow are mildly or sub-relativistic with $\beta_s \sim 0.01 - 0.4$ and have $M_{s}\lesssim 2$. Shocks in the shocked ICM are non-relativistic and weak, and dynamically not very important. The dissipation rate of the kinetic energy at shocks, normalized to the jet energy injection rate, $\mathcal{L}_{\rm shock,tot}/Q_j$, is higher for lower $Q_j$, while $\mathcal{L}_{\rm shock,tot}$ itself is larger for higher $Q_j$. The backflow is the most important in the shock dissipation for the Q45 and Q46 models, while both the backflow and jet spine flow are about equally important for the Q47 models. We have found that $\mathcal{L}_{\rm shock,tot}/Q_j$ is $\sim 0.5 - 1$, $0.45-0.8$, and $0.1-0.15$ around the end of our simulations for the Q45, Q46, and Q47 models, respectively. We expect that the DSA of CRs would be important in radio galaxies, since a substantial fraction of the jet energy is dissipated at shocks.

{\it Velocity shear:} We have examined the strength of velocity shear, $\Omega_{\rm shear}=|{\partial v_z / \partial r}|$, and the relativistic shear coefficient, $S_{r}$, in Equation (\ref{relshearcoeff}). The PDF of $\Omega_{\rm shear}$, $N(\Omega_{\rm shear})$, does not strongly depend on $Q_j$, and peaks at $\Omega_{\rm shear}(r_j/c)\gtrsim 1$ for the jet spine flow, $\sim 0.1-1$ for the backflow, and $\lesssim 0.01$ for the shocked ICM, respectively. The PDF, $S_{r}$, $N(\mathcal{S}_{r})$ in the jet spine flow extends up to $\mathcal{S}_{r} (r_j/c)^2\sim 10^3-10^5 $, depending on $Q_j$, but peaks at $\sim 0.1-0.2$, almost independent of $Q_j$. In the backflow, $N(\mathcal{S}_{r})$ peaks at $\mathcal{S}_{r} (r_j/c)^2 \sim 10^{-3} - 10^{-2}$. With a large volume of the cocoon, the production of CRs via the gradual shear acceleration would be the most efficient in the backflow \citep[e.g.,][]{rieger2004}, while the further energization to higher energies could proceed through the discrete shear acceleration at the interface of the jet-backflow \citep{kimura2018}.

{\it Vorticity:} As a measure of turbulence, we have quantified the total vorticity, $\Omega_t = |\mbox{\boldmath$\nabla$}\times\mbox{\boldmath$v$}|$, and the vorticity excluding the shear contribution, $\Omega_- = |\mbox{\boldmath$\nabla$}\times\mbox{\boldmath$v$}+{\partial v_{z}}/{\partial r}\hat{\mbox{\boldmath$\theta$}}|$. The PDFs of $\Omega_t$ and $\Omega_-$ are similar to that of $\Omega_{\rm shear}$, and they do not strongly depend on $Q_j$.  The fact that $\Omega_- \gtrsim \Omega_{\rm shear}$ indicates that along with the shear acceleration, the stochastic turbulent acceleration of CRs could be also important in radio galaxies. In reality, both the shear and turbulent accelerations may work together and their interplay could be important \citep[e.g.,][]{ohira2013}.

{\it Turbulence.} The turbulence generated in the jet-induced structures shows the Kolmogorov spectrum, $P_v(k)\propto k^{-5/3}$, agreeing with the results of previous studies of RHD turbulence \citep[e.g.,][]{zrake2013,radice2013}. Assuming the Kolmogorov scaling for turbulent cascade, we have estimated the turbulent dissipation rate of the jet kinetic energy, normalized to the jet power, $\mathcal{L}_{\rm turb}/Q_j$. As for the shock dissipation, the normalized dissipation rate, $\mathcal{L}_{\rm turb,tot}/Q_j$, is higher for lower $Q_j$, while $\mathcal{L}_{\rm turb,tot}$ itself is larger for higher $Q_j$. An interesting point is that $\mathcal{L}_{\rm turb,tot}$ is determined mainly by $Q_j$, while it is almost independent of $\eta$ and $\zeta$. The turbulent dissipation is the greatest in the backflow for the Q45 and Q46 models. For the high-power Q47 models, all the contributions from the jet spine flow, backflow, and shocked ICM seem to be important. We have found that $\mathcal{L}_{\rm turb,tot}/Q_j$ is $\sim 0.65$, $0.25$, and $0.07$ for the Q45, Q46, and Q47 models, respectively. In our estimations, $\mathcal{L}_{\rm shock,tot}$ is somewhat larger than $\mathcal{L}_{\rm turb,tot}$. But considering the uncertainties involved in the estimations, we suggest that the two types of dissipation would be comparable, and hence both shocks and turbulence/shear could be important in the acceleration of CRs in radio galaxies.

In upcoming papers, using the results of the jet simulations obtained in this study, we aim to present quantitative studies for the acceleration of UHECRs by different mechanisms such as DSA, shear acceleration, and stochastic turbulent acceleration, in FR-II radio galaxies.

\begin{acknowledgments}
This work was supported by the National Research Foundation (NRF) of Korea through grants 2016R1A5A1013277, 2020R1A2C2102800, and 2020R1F1A1048189. The work of J.S. was also supported by the NRF through grant 2020R1A6A3A13071702. Some of simulations were performed using the high performance computing resources of the UNIST Supercomputing Center.
\end{acknowledgments}

\bibliography{RadioGalaxy}{}
\bibliographystyle{aasjournal}

\end{document}